\documentclass{aa}
\usepackage{times}
\usepackage{graphics}

\begin{document}

\thesaurus{08(08.06.2; 09.03.1; 09.13.2; 13.19.3)}

\title{HNCO in massive galactic dense cores
\thanks{Based on the observations collected at the European Southern
Observatory, La Silla, Chile and
on observations with the Heinrich-Hertz-Telescope (HHT). 
The HHT is operated by the Submillimeter Telescope Observatory on
behalf of Steward Observatory and the MPI f\"ur Radioastronomie.}
\thanks{Tables~1, 2, 5, 6 are also available in electronic form and
Tables~7--14 are only available in electronic form at the 
CDS via anonymous ftp to cdsarc.u-strasbg.fr (130.79.128.5) or via
http://cdsweb.u-strasbg.fr/Abstract.html}
}

\author{I.~Zinchenko\inst{1,2}\and
	C.~Henkel\inst{3} \and
	R.Q.~Mao\inst{3,4} 
}

\offprints{I.~Zinchenko}
\mail{zin@appl.sci-nnov.ru}

\institute{Institute of Applied Physics of the Russian Academy of Sciences,
	   46 Ulyanov~str., 603600 Nizhny Novgorod, Russia
\and Helsinki University Observatory, T\"ahtitorninm\"aki, P.O.Box 14,
	   FIN-00014 University of Helsinki, Finland
\and Max-Planck-Institut f\"ur Radioastronomie, Auf dem H\"ugel 69,
Bonn, Germany
\and Purple Mountain Observatory, 210008 Nanjing, China
}

\date{Received date  ; accepted date  }

\maketitle

\begin{abstract}

We surveyed 81 dense molecular cores associated with regions 
of massive star formation and Sgr A in the $J_{\rm K_{-1}K{_1}} = 
5_{05}-4_{04}$ and $10_{010}-9_{09}$ lines of HNCO. Line emission was
detected towards 57 objects. Selected subsamples were also
observed in the $1_{01}-0_{00}$, $4_{04}-3_{03}$, $7_{07}-6_{06}$, 
$15_{015}-14_{014}$, $16_{016}-15_{015}$ and $21_{021}-20_{020}$ lines, 
covering
a frequency range from 22 to 461 GHz. HNCO lines from the 
$K_{-1} = 2,3$ ladders were detected in several sources. Towards
Orion-KL, $K_{-1} = 5$ transitions with upper state energies
$E_{\rm u}/k \sim 1100$ and 1300 K could be observed.

Five HNCO cores were mapped. The sources remain spatially unresolved 
at 220 and 461 GHz ($10_{010}-9_{09}$ and $21_{010}-20_{020}$
transitions) with beam sizes of 24\arcsec\ and 18\arcsec, respectively.

The detection of hyperfine structure in the $1_{01}-0_{00}$ transition 
is consistent with optically thin emission under conditions of Local
Thermodynamic Equilibrium (LTE). This is corroborated by a rotational 
diagram analysis of Orion-KL
that indicates optically thin line emission also for
transitions between higher excited states.
At the same time a tentative detection of interstellar
HN$^{13}$CO  (the $10_{0,10}-9_{0,9}$ line at 220 GHz toward G~310.12-0.20)
suggests optically thick emission from some rotational transitions.

Typical HNCO abundances relative to H$_2$ as 
derived from a population diagram
analysis are $\sim 10^{-9}$. The rotational temperatures reach $\sim 500$~K.
The gas densities in regions of HNCO $K_{-1}=0$ emission should be
$n\ga 10^6$~cm$^{-3}$ and in regions of $K_{-1}>0$ emission about an order
of magnitude higher even for radiative excitation.

HNCO abundances are found to be enhanced in high-velocity gas. HNCO integrated
line intensities correlate well with those of thermal SiO emission. This
indicates a spatial coexistence of the two species and may hint at a common 
production mechanism, presumably based on shock chemistry.

\keywords{Stars: formation -- ISM: clouds --
ISM: molecules -- Radio lines: interstellar}
\end{abstract}

\section{Introduction}

Systematic studies of dense molecular
cores in regions of high mass star formation (HMSF) are of great
importance for our general understanding of star formation.
In comparison with low mass
star formation regions, so far only a few rather arbitrarily selected cores
associated with HMSF have been investigated in some detail. 

In recent years we performed extensive surveys of dense cores in regions of
high mass star formation, mainly in CS (Zinchenko et al. 1995,
1998). We used water masers as signposts of high mass star formation.
Both outer and inner Galaxy were covered by these surveys 
($l\approx 120\degr-210\degr$ and $260\degr-308\degr$).
The innermost part of the Galaxy ($l\approx 308\degr-360\degr$)
was observed in a similar way by
Juvela (1996). In addition, sources associated with 
water masers were surveyed in thermal SiO
(\cite{harju98}) which is supposed to be a good indicator of
shocks in molecular clouds. From these observations we derived basic
physical parameters of the cores and constructed their statistical
distributions (\cite{zin95}, \cite{zin98}). 
In order to investigate a range of core
densities, observations of lines with different excitation conditions
are needed. One of the interesting candidates is the HNCO (isocyanic
acid) molecule.

HNCO was first detected by Snyder \& Buhl (1972) in Sgr~B2. 
Subsequent studies have concentrated mostly on the Galactic center region
where the HNCO emission was found to be particularly strong 
(e.g., Churchwell et al. 1986, Wilson et al. 1996, Lindqvist et al. 1995,
Kuan \& Snyder 1996, Dahmen et al. 1997, Sato et al. 1997). A
survey of HNCO emission throughout the Galaxy was made by Jackson et al. 
(1984) in the $J_{\rm K_{-1}K{_1}} = 5_{05}-4_{04}$ and $4_{04}-3_{03}$ 
transitions with the 11~m NRAO telescope. Seven (from 18) clouds including 
Orion~KL 
were detected at rather low levels of intensity
(typically $\sim 0.2$~K on a $T_{\rm A}^*$ scale).
Churchwell et al. (1986) obtained strict upper limits on HNCO $1_{01}-0_{00}$
and $2_{02}-1_{01}$ emission towards about 20 galactic sources with the 36.6~m
Haystack antenna.

HNCO is a slightly asymmetric rotor. Its levels may be designated as
$J_{\rm K_{-1}K_1}$ where $J$ is the total angular momentum and
$K_{-1}$, $K_1$ are quantum numbers corresponding to the projection of
$J$ on the symmetry axis for the limiting
cases of prolate and oblate symmetric top, respectively (e.g. Townes
\& Schawlow 1975). The structure of the HNCO energy levels can be
represented as a set of ``ladders'' with different $K_{-1}$ values,
like for a symmetric top. However, due to the asymmetry of the
molecule radiative transitions between different $K_{-1}$ ladders
($b$-type transitions) are allowed and, moreover, they are very
fast. The corresponding component of the dipole moment is similar to its
component for transitions inside the $K_{-1}$ ladders ($a$-type transitions).
Churchwell et al. (1986) found that as a result the HNCO
excitation is governed mostly by radiative rather than collisional
processes (at least in Sgr~B2).

On the basis of their estimates of source parameters Jackson et al. (1984)
concluded that HNCO is a potentially valuable probe of the densest regions
($n\ga 10^6$~cm$^{-3}$) of molecular clouds. It was shown also that HNCO
is rather sensitive to far infrared (FIR) radiation fields 
due to the fact that the
lowest levels of the $K_{-1}=0,$ 1 and $K_{-1}=1,$ 2 ladders are 
separated by energies corresponding to FIR wavelengths (330~$\mu$m and
110~$\mu$m, respectively). 

From this consideration it is clear that multitransitional data are needed
to understand HNCO excitation and to derive the source properties.
Bearing this in mind we undertook a survey of HNCO emission 
in various rotational lines, also trying to detect emission from higher
excited $K-$ladders ($K_{-1}>0$).
Five cores were mapped in HNCO to estimate the extent of the emission.

Several other species were observed simultaneously with HNCO. The
most prominent are C$^{18}$O and SO. In the following we thus also
compare HNCO with C$^{18}$O.

\section{Observations} \label{sec:obs}

\subsection{Source list}

For this study we observed those dense cores showing particularly
strong CS emission ($T_{\rm mb}> 3$~K) in the surveys of Zinchenko et al. 
(1995, 1998)
and Juvela (1996). Several strong SiO ($v=0$) sources detected
by Harju et al. (1998) are also included in our sample. 
Sources
observed at the SEST and at Onsala are presented in Tables~\ref{table:sest},
\ref{table:oso}. Sources also observed at Effelsberg or at the HHT are marked 
in both tables. 

We designate most sources according to their galactic coordinates. 
Exceptions are Orion~KL and Sgr~A. For Sgr~A we use the position
observed by Jackson et al. (1984) for comparison (known as the M-0.13-0.08
cloud, see Lindqvist et al. 1995).
Common identifications with some well known objects are given in the last 
column.

\begin{table}
\caption[]{Source list for SEST observations.}
\begin{tabular}{llrl}
\hline\noalign{\smallskip}
Name
 &$\alpha(1950)$ &$\delta(1950)$ &Remarks\\
&($^{\rm h}$) ($^{\rm m}$) ($^{\rm s}$)
&($^\circ$) ($^\prime$) ($^{\prime\prime}$)\\
\noalign{\smallskip}\hline\noalign{\smallskip}
G 261.64$-$2.09    &08 30 23.2    &$-$43 03 31 \\
G 264.28$+$1.48    &08 54 39.0    &$-$42 53 30     &RCW 34\\
G 265.14$+$1.45    &08 57 36.3    &$-$43 33 38     &RCW 36\\
G 267.94$-$1.06    &08 57 21.7    &$-$47 19 04     &RCW 38\\
G 268.42$-$0.85    &09 00 12.1    &$-$47 32 07  \\
G 269.16$-$1.14    &09 01 51.6    &$-$48 16 43  \\
G 270.26$+$0.83    &09 14 58.0    &$-$47 44 00     &RCW 41\\
G 285.26$-$0.05    &10 29 36.8    &$-$57 46 40 \\
G 286.20$+$0.17    &10 36 34.8    &$-$58 03 22  \\
G 291.27$-$0.71    &11 09 42.0    &$-$61 01 55 \\
G 291.57$-$0.43    &11 12 54.0    &$-$60 52 57   &NGC 3603\\
G 294.97$-$1.73    &11 36 51.6    &$-$63 12 09 \\
G 300.97$+$1.14    &12 32 00.2    &$-$61 23 44     &RCW 65\\
G 301.12$-$0.20    &12 32 31.3    &$-$62 44 38  \\
G 305.20$+$0.21    &13 07 59.9    &$-$62 18 50     &RCW 74\\
G 305.36$+$0.21    &13 09 21.2    &$-$62 18 02   \\
G 308.80$-$0.25    &13 13 27.2    &$-$62 42 56 \\
G 308.00$+$2.02    &13 29 24.3    &$-$60 11 22   \\
G 308.92$+$0.12    &13 39 34.4    &$-$61 53 45    &RCW 79\\
G 309.92$+$0.48    &13 47 12.5    &$-$61 19 58  \\
G 316.77$-$0.02    &14 41 10.4    &$-$59 35 30  \\
G 316.81$-$0.06    &14 41 36.4    &$-$59 36 53 \\
G 318.05$+$0.09    &14 49 51.9    &$-$58 56 40  \\
G 323.74$-$0.25    &15 27 49.8    &$-$56 20 15 \\
G 324.20$+$0.12    &15 29 01.2    &$-$55 46 12 \\
G 326.47$+$0.70    &15 39 28.2    &$-$53 58 01  \\
G 326.64$+$0.61    &15 40 42.6    &$-$53 56 29 \\
G 328.30$+$0.43    &15 50 15.3    &$-$53 02 46 \\
G 328.81$+$0.63    &15 51 59.0    &$-$52 34 24  \\
G 328.24$-$0.54    &15 54 04.9    &$-$53 50 09   \\
G 329.03$-$0.20    &15 56 40.1    &$-$53 04 08 \\
G 330.95$-$0.19    &16 06 03.4    &$-$51 47 30 \\
G 330.88$-$0.36    &16 06 30.0    &$-$51 58 14  \\
G 331.28$-$0.18    &16 07 36.0    &$-$51 33 40  \\
G 332.83$-$0.55    &16 16 23.7    &$-$50 45 45     &RCW 106\\
G 333.13$-$0.43    &16 17 12.6    &$-$50 28 18  \\
G 333.60$-$0.22    &16 18 24.5    &$-$49 59 08  \\
G 337.40$-$0.40    &16 35 08.1    &$-$47 22 23  \\
G 340.06$-$0.25    &16 44 36.4    &$-$45 16 26    \\
G 345.01$+$1.80    &16 53 18.8    &$-$40 09 36     \\
G 343.12$-$0.06    &16 54 42.8    &$-$42 47 49  \\
G 345.51$+$0.35    &17 00 53.6    &$-$40 40 02   \\
G 345.00$-$0.23    &17 01 40.7    &$-$41 25 07  \\
G 345.41$-$0.94    &17 06 02.3    &$-$41 31 44  \\
G 348.55$-$0.97    &17 15 53.1    &$-$39 00 57  \\
G 350.10$+$0.09    &17 16 01.0    &$-$37 07 30     \\
G 348.73$-$1.04    &17 16 39.7    &$-$38 54 17    &RCW 122\\
G 351.41$+$0.64    &17 17 32.5    &$-$35 44 13   \\
G 351.58$-$0.36    &17 22 04.2    &$-$36 10 11   \\  
G 351.78$-$0.54    &17 23 20.9    &$-$36 06 53    \\
G 353.41$-$0.36    &17 27 06.5    &$-$34 39 41   \\
G 359.97$-$0.46    &17 44 10.4    &$-$29 11 03   \\
Orion KL$^{a,b}$          &05 32 47.0    &$-$05 24 23 \\
G 173.48$+$2.45$^c$	&05 35 51.3 &35 44 16 &S231\\
G 192.60$-$0.05$^b$ &06 09 58.2  &18 00 17  &S255\\
Sgr A$^a$              &17 42 28.0    &$-$29 04 01 \\
\noalign{\smallskip}\hline\noalign{\smallskip}
\end{tabular}

$^a$ also observed in Effelsberg, $^b$ also observed with HHT,
$^c$ also observed in Onsala.
\label{table:sest}
\end{table}

\begin{table}
\caption[]{Source list for Onsala observations.}
\begin{tabular}{llll}
\hline\noalign{\smallskip}
Name
 &$\alpha(1950)$ &$\delta(1950)$ &Remarks\\
&($^{\rm h}$) ($^{\rm m}$) ($^{\rm s}$)
&($^\circ$) ($^\prime$) ($^{\prime\prime}$)\\
\noalign{\smallskip}\hline\noalign{\smallskip}
G 121.30$+$0.66	&00 33 53.3 &63 12 32 &RNO 1B\\
G 123.07$-$6.31	&00 49 29.2 &56 17 36 &NGC 281\\
G 133.69$+$1.22	&02 21 40.8 &61 53 26 &W3 (1)\\
G 133.95$+$1.07$^b$&02 23 17.3 &61 38 58 &W3 (OH)\\
G 170.66$-$0.27 &05 16 53.6  &36 34 21 &IRAS05168+3634   \\
G 173.17$+$2.35   &05 34 35.9  &35 56 57 &IRAS05345+3556    \\
G 173.48$+$2.45$^{b,c}$	&05 35 51.3 &35 44 16 &S 231\\
G 173.72$+$2.70	&05 37 31.8 &35 40 18 &S 235\\
G 188.95$+$0.89   &06 05 53.7 &21 39 09 &S 247\\
G 34.26$+$0.15    &18 50 46.4  &01 11 10 &IRAS18507+0110    \\ 
G 40.50$+$2.54    &18 53 47.0  &07 49 26 &S76 E  \\ 
G 43.17$+$0.01$^{a,b}$    &19 07 49.8  &09 01 17 &W49 N   \\ 
G 49.49$-$0.39$^{a,b}$    &19 21 26.2  &14 24 44 &W51 M   \\ 
G 60.89$-$0.13  &19 44 14.0  &24 28 10 &S87         \\ 
G 61.48$+$0.10    &19 44 42.0  &25 05 30 &S88B        \\ 
G 70.29$+$1.60    &19 59 50.0  &33 24 17 &K3-50  \\ 
G 69.54$-$0.98$^b$  &20 08 09.9  &31 22 42 &ON1    \\ 
G 77.47$+$1.77    &20 18 50.0  &39 28 45 &JC20188+3928 \\
G 75.78$+$0.34      &20 19 51.8  &37 17 01 &ON2 N \\
G 81.87$+$0.78$^b$      &20 36 50.5  &42 27 01 &W75 N \\ 
G 81.72$+$0.57$^{a,b}$      &20 37 13.7  &42 12 11 &W75 (OH) \\ 
G 81.77$+$0.60      &20 37 16.6  &42 15 15 &W75 S3 \\ 
G 92.67$+$3.07    &21 07 46.7  &52 10 23 &J21078+5211  \\
G 99.98$+$4.17    &21 39 10.3  &58 02 29 &IRAS21391+5802\\
G 108.76$-$0.95 &22 56 38.4  &58 31 04 &JC22566+5830 \\
G 108.76$-$0.98 &22 56 45.2  &58 29 10 &S152(OH)    \\
G 111.53$+$0.76$^{a,b}$   &23 11 36.1  &61 10 30 &S 158 \\ 
\noalign{\smallskip}\hline\noalign{\smallskip}
\end{tabular}

$^a$ also observed in Effelsberg, $^b$ also observed with HHT,
$^c$ also observed with SEST.
\label{table:oso}
\end{table}

\subsection{Observational procedures}

The most important parameters of our SEST-15m, OSO-20m, Effelsberg 100-m
and HHT measurements are summarized in Tables~\ref{table:obs},
\ref{table:freq}. 
Further details are given below for each instrument.

\begin{table*}
\caption[]{Observing parameters 
($\Delta\nu$ is the spectral resolution).
}
\begin{tabular}{llrllllll}
\hline\noalign{\smallskip}
Molecule &Transition$^a$ &Frequency &Telescope &Date &HPBW &$\eta_{\rm mb}$
&$T_{\rm sys}^b$ &$\Delta\nu$\\
&&(MHz) &&&(\arcsec) &&(K) &(kHz)\\
\noalign{\smallskip}\hline\noalign{\smallskip}
HNCO	&$1_{01}-0_{00}$ &21981.460 &Eff. 100m&1998	&40 &0.3 &50--100	&12.5\\
	&$4_{04}-3_{03}$ &87925.252 &OSO 20m	&1997	&40$^c$ &0.60$^c$ &210--290 &250\\
	&$5_{05}-4_{04}$ &109905.758 &OSO 20m	&1997	&35$^c$ &0.52$^c$ &300--450 &250\\
	&$5_{05}-4_{04}$ &109905.758 &SEST 15m	&1997	&47$^c$ &0.71$^c$ &200--270 &86\\
	&$7_{07}-6_{06}$ &153865.080 &SEST 15m	&1997	&33$^c$ &0.64$^c$ &150--180 &86\\
	&$10_{0,10}-9_{0,9}$ &219798.320 &SEST 15m&1997	&24$^c$ &0.52$^c$ &190--360 &86\\
	&$15_{0,15}-14_{0,14}$ &329664.535 &HHT 10m&1999	&25 &0.50 &900--6000 &480\\
	&$16_{0,15}-15_{0,14}$ &351633.457 &HHT 10m&1999	&24 &0.50 &700--1000 &480\\
	&$21_{0,21}-20_{0,20}$ &461450.670 &HHT 10m&1999	&18 &0.38 &900--2000 &480\\
C$^{18}$O&$1-0$	&109782.160 &OSO 20m	&1997	&35$^c$ &0.52$^c$&300--450 	&1000\\
	&$2-1$	&219560.319 &SEST 15m	&1997	&24$^c$ &0.52$^c$&190--360	&1400\\
\noalign{\smallskip}\hline\noalign{\smallskip}
\end{tabular}

$^a$The frequencies and spectral resolutions for the observed
$K_{-1}>0$ transitions are presented in Table~\ref{table:freq}.\\
$^b$The system temperatures are given on a $T_{\rm A}^*$ scale.\\
$^c$Beam sizes and main beam efficiencies are obtained by interpolating
the data from the SEST manual (for SEST) and those provided by L.E.B.~Johansson
(for OSO) at nearby frequencies.
\label{table:obs}
\end{table*}

\begin{table}
\caption[]{Frequencies and spectral resolutions for the observed 
$K_{-1}>0$ transitions. For other observing parameters see
Table~\ref{table:obs}}
\begin{tabular}{lll}
\hline\noalign{\smallskip}
Transition &Frequency &$\Delta\nu$\\
&(MHz) &(kHz)\\
\noalign{\smallskip}\hline\noalign{\smallskip}
	$10_{2,9}-9_{2,8}$ &219733.850 &1400\\
	$10_{2,8}-9_{2,7}$ &219737.193 &1400\\
	$10_{3,8}-9_{3,7}$ &219656.710 &1400\\
	$10_{3,7}-9_{3,6}$ &219656.710 &1400\\
	$10_{4,7}-9_{4,6}$ &219547.082 &1400\\
	$10_{4,6}-9_{4,5}$ &219547.095 &1400\\
	$10_{5,6}-9_{5,5}$ &219392.412 &1400\\
	$10_{5,5}-9_{5,4}$ &219392.412 &1400\\
	$15_{2,14}-14_{2,13}$ &329573.46 &480\\
	$15_{2,13}-14_{2,12}$ &329585.09 &480\\
	$21_{2,20}-20_{2,19}$ &461336.93 &480\\
	$21_{2,19}-20_{2,18}$ &461368.88 &480\\
	$21_{3,18}-20_{3,17}$ &461182.51 &480\\
	$21_{3,19}-20_{3,18}$ &461182.45 &480\\
	$21_{4,17}-20_{4,16}$ &460950.89 &480\\
	$21_{4,18}-20_{4,17}$ &460950.89 &480\\
	$21_{5,16}-20_{5,15}$ &460625.75 &480\\
	$21_{5,17}-20_{5,16}$ &460625.75 &480\\
\noalign{\smallskip}\hline\noalign{\smallskip}
\end{tabular}
\label{table:freq}
\end{table}

\subsubsection{SEST observations}

\begin{figure}
\resizebox{\hsize}{!}{\rotatebox{-90}{\includegraphics{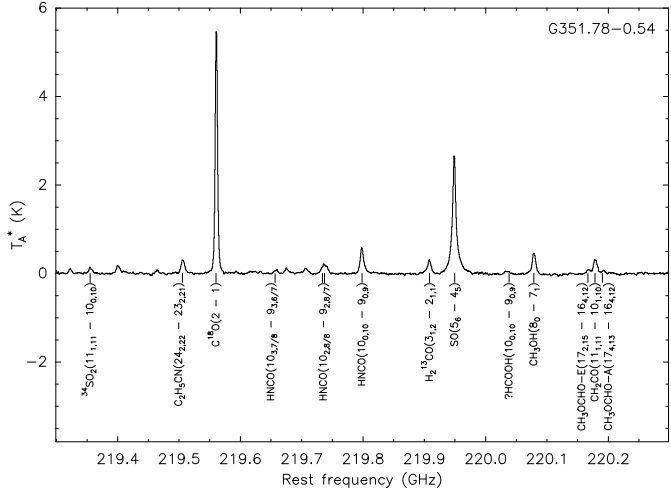}}}
\caption[]{A SEST low resolution spectrum. For identified features,
molecular species and transitions are given}
\label{fig:sest-lr}
\end{figure}

The observations were performed with SIS receivers in a single-sideband (SSB) 
mode using  dual beam switching with a
beam throw of $\sim 12$\arcmin. 
At 220~GHz we used 2 acousto-optical spectrometers in parallel:
(1) a 2000 channel high-resolution 
spectrometer (HRS) with 86~MHz bandwidth, 43 kHz channel separation and
80~kHz resolution and (2) a 1440 channel low-resolution spectrometer (LR1) 
with a 1000~MHz total bandwidth, 0.7~MHz channel separation and 1.4~MHz
spectral resolution. The LR1 band was centered on the HNCO $10_{0,10}-9_{0,9}$
transition. However, it covered some other HNCO transitions too (see 
Table~\ref{table:freq}) as well as C$^{18}$O (2--1), SO ($6_5-5_4$) and
other lines (Fig.~\ref{fig:sest-lr} shows a typical spectrum).

The 110 and 154~GHz observations were performed 
simultaneously; the spectra were recorded by the HRS which band was split
into two equal parts. The 220~GHz HRS spectra were smoothed to 170~kHz
resolution and the 110 and 154~mm spectra were smoothed to 86~kHz resolution.
Pointing was checked periodically by observations of nearby SiO
masers; the pointing accuracy was $\la 5$\arcsec.

The standard chopper-wheel technique was used for the calibration. 
We express
the results in units of main beam brightness temperature ($T_{\rm
mb}$) assuming the main beam efficiencies ($\eta_{\rm mb}$) as given in
Table~\ref{table:obs}.
The temperature scale
was checked by observations of Orion~KL.

In most sources only one position was observed, corresponding typically
to the peak of the CS emission. In addition, G~270.26+0.83 and G~301.12$-$0.20
were mapped with a spacing of 10\arcsec. 

\subsubsection{Onsala observations}

At Onsala, the 110\,GHz observing procedure was 
very similar to that at the SEST.
The observations were also performed in a dual beam switching mode with a
beam throw of 11{\farcm}5. The front-end was a SIS receiver tuned to SSB
operation. As backend we used 2 filter spectrometers
in parallel: a 256 channel filterbank with 250~kHz resolution and a 512 channel
filterbank with 1~MHz resolution. The calibration procedure was the same as
at the SEST. The pointing accuracy checked by observations of nearby SiO
masers was $\la 5$\arcsec. The strongest HNCO source from the Onsala sample,
W51M, was mapped with 40\arcsec\ spacing.

\subsubsection{Effelsberg observations}

The 22\,GHz observations in Effelsberg were performed with a K-band maser
amplifier using position switching. The offset positions were 
displaced by 10\arcmin--15\arcmin\ symmetrically in azimuth.
Pointing was checked periodically by observations of nearby continuum
sources; the pointing accuracy was $\la 10$\arcsec. The integration
time per position was a few hours.

The main beam temperature scale was checked by observations of nearby 
continuum calibration sources, NGC~7027 and W3(OH); for Sgr~A we 
used Sgr~B2.
The fluxes for the first two sources were taken from Ott et al. (1994). 
The Sgr~B2
flux at 1.3~cm was taken from Mart\'{\i}n-Pintado et al. (1990).

\subsubsection{HHT observations}

To observe the HNCO $J$ = 21--20 lines at 461\,GHz we have used 
the Heinrich Hertz Telescope (HHT) 
on Mt. Graham (Baars \& Martin 1996) during Feb. 1999 with a
beamwidth of 18$''$. Spectra were taken employing an SIS receiver with
backends consisting of two acousto optical spectrometers with 2048 channels
each, channel spacing $\sim$480 and $\sim$120\,kHz, frequency resolution
$\sim$930 and 230\,kHz, and total bandwidths of $\sim$1\,GHz and 250\,MHz,
respectively. Receiver temperatures were $\sim$150\,K, system temperatures
were $\sim$1000\,K on a $T_{\rm A}^{*}$ scale. The receiver was sensitive to
both sidebands. Any imbalance in the gains of the lower and upper sideband
would thus lead to calibration errors. To account for this, we have observed
the CO $J$ = 4--3 line of Orion-KL with the same receiver tuning setup and
obtain $T_{\rm A}^{*}$ $\sim$ 70\,K, in good agreement with Schulz et al.
(1995). 

HNCO $J=16-15$ (351.63346 GHz) and $J=15-14$ (329.66454 GHz) line emission 
was observed with a dual channel
SIS receiver in early April 1999 at the HHT. The beamwidth was 22\arcsec, 
receiver temperatures were 135 K;
system temperatures were $\sim$700 K on a $T_{\rm A}^*$ scale.
The receivers were also sensitive to both sidebands. 
We have used published spectra from Orion-KL and IRC+10216 as
calibrators (\cite{groesbeck94,schilke97}).

All results displayed are given in units of main beam brightness temperature
($T_{\rm mb}$). This is related to $T_{\rm A}^{*}$ via $T_{\rm mb}$ = 
$T_{\rm A}^{*}$ ($F_{\rm eff}$/$B_{\rm eff}$) (cf. Downes 1989). The main
beam efficiency, $B_{\rm eff}$, was 0.38 at 461 GHz and 0.5 at 330 and 352
GHz as obtained by measurements of 
Saturn. The forward hemisphere efficiency, $F_{\rm eff}$, is 0.75 at 461 GHz
and 0.9 at 330 and 352 GHz
(D. Muders, priv. comm.). The HHT is with an rms surface deviation of 
$\sim$20$\mu$m (i.e. $\lambda$/30 at 461\,GHz) quite accurate. 
Thus emission from the sidelobes should not be a problem.

Pointing was obtained toward Jupiter (continuum pointing) and toward Orion-KL
and R Cas (line pointing) with maximum deviations of order 5$''$. Observations
were carried out in a position switching mode with the off-position
$\sim$1000$''$ offset from the source position.

\subsection{Data reduction and analysis}

We have reduced the data and produced maps using the  GAG
({\it Groupe d'Astrophysique de Grenoble}) software package.
The measured spectra were fitted by one or more
gaussian components. 

\section{Results} \label{sec:results}

\subsection{One-point observations}

HNCO was detected in 36 SEST sources (from 56 observed) and in 22 OSO sources
(from 27). Because of one source belonging to both samples, the total number
of detected objects is 57. 
In many cases $K_{-1}>0$ transitions were detected too.
The gaussian line parameters are presented 
in Tables~5--14 (Tables~7--14 are available only electronically).
It is worth noting that a single-gaussian fit is clearly insufficient in
many cases because the lines have broad wings and other non-gaussian
features. Therefore, the values in the tables give only a rough
representation of the line profiles
(the integrated intensities were obtained by integrating over the lines
in most cases).

Table~\ref{table:1.4mm} summarizes the 220~GHz SEST results for HNCO 
$10_{0,10}-9_{0,9}$ and C$^{18}$O. 
The Onsala $5_{05}-4_{04}$ and C$^{18}$O 
results are presented in Table~\ref{table:oso-rst}.
The 220~GHz results for  the 
$K_{-1}=2$, 3 ladders are given in Tables~7, 8.
The 110 and 154~GHz SEST data are displayed in Table~9.
The Onsala 88~GHz data are summarized in Table~10.
The Effelsberg data are presented in Table~11. Tables~12--14 contain 
the HHT data.
We fitted the
Effelsberg spectra with 3-component gaussians with fixed separations
corresponding to the hyperfine structure of the $1_{01}-0_{00}$ transition.

\begin{table*}
\caption[]{C$^{18}$O ($2-1$) and HNCO ($10_{0,10}-9_{0,9}$) integrated
intensities 
and gaussian line parameters at the indicated positions (cf. 
Tables~\ref{table:sest},\ref{table:oso}) 
measured at SEST.
The numbers in the brackets are the statistical
uncertainties in the last digits (standard deviations).}
\begin{flushleft}
\begin{tabular}{lrrlrrlllrrl}\hline\noalign{\smallskip}
&&&\multicolumn{4}{l}{C$^{18}$O ($2-1$)}
&&\multicolumn{4}{l}{HNCO ($10_{0,10}-9_{0,9}$)}
\\
\cline{4-7}
\cline{9-12}
\noalign{\smallskip}
Source
&$\Delta \alpha$  &$\Delta \delta$
&$\int T_{\rm mb} dv$
&\multicolumn{1}{l}{$T_{\rm mb}$}
&\multicolumn{1}{l}{$V_{\rm LSR}$}
&$\Delta V$
&
&$\int T_{\rm mb} dv$
&\multicolumn{1}{l}{$T_{\rm mb}$}
&\multicolumn{1}{l}{$V_{\rm LSR}$}
&$\Delta V$
\\
&(\arcsec) &(\arcsec)
&(K$\cdot$km/s)
&\multicolumn{1}{l}{(K)} 
&\multicolumn{1}{l}{(km/s)}
&\multicolumn{1}{l}{(km/s)}
&
&(K$\cdot$km/s)
&\multicolumn{1}{l}{(K)} 
&\multicolumn{1}{l}{(km/s)}
&\multicolumn{1}{l}{(km/s)}
\\
\noalign{\smallskip}
\hline\noalign{\smallskip}
G 261.64     &   20&    0&  17.22(05)&   3.73(01)&  13.76(01)&   4.02(02)&&   0.63(05)&   0.15(01)&  13.69(16)&   3.87(39)\\
G 264.28     &    0&  --40&   3.41(04)&   0.90(01)&   5.39(02)&   3.36(05)&&   0.17(03)&   0.07(01)&   6.12(25)&   2.32(46)\\
G 265.14     &  --40&    0&  19.16(06)&   4.63(02)&   6.75(01)&   3.68(02)\\
G 267.94     &    0&    0&  14.08(07)&   2.59(01)&   1.04(01)&   4.68(03)\\
G 268.42     &    0&    0&  29.38(06)&   5.57(01)&   2.55(00)&   4.60(01)&&   0.28(05)&   0.07(01)&   3.01(36)&   3.92(73)\\
G 269.16     &    0&   40&  24.95(08)&   4.30(02)&   9.72(00)&   5.09(02)&&   0.79(07)&   0.18(02)&   9.64(18)&   4.05(44)\\
G 270.26     &  --20&   20&  16.28(05)&   3.24(01)&   8.67(01)&   4.55(02)&&   1.07(06)&   0.24(02)&   9.92(12)&   4.25(30)\\
G 285.26     &    0&    0&  10.16(06)&   1.76(01)&   2.47(01)&   5.11(04)&&   0.50(09)&   0.07(01)&   1.64(65)&   7.06(156)\\
G 286.20     &   40&  --40&  14.53(04)&   3.14(01)& --20.86(01)&   4.14(01)\\
G 291.27     &  --40&  --40&  28.39(06)&   5.21(01)& --23.95(01)&   4.93(01)&&   0.56(07)&   0.12(02)& --23.80(25)&   4.25(79)\\
G 291.57     &    0&    0&   7.59(07)&   1.13(01)&  13.12(03)&   6.03(07)\\
G 294.97     &    0&    0&  15.77(05)&   4.02(02)&  --9.24(01)&   3.31(01)\\
G 300.97     &    0&   40&  23.13(05)&   4.39(01)& --43.89(01)&   4.64(01)\\
G 301.12     &   80&  --80&  44.35(18)&   7.13(03)& --40.01(01)&   5.44(03)&&   4.15(08)&   0.61(01)& --39.37(06)&   6.43(16)\\
G 305.20     &    0&    0&  22.98(07)&   3.41(01)& --42.13(01)&   6.39(02)&&   1.84(10)&   0.17(01)& --40.73(26)&  10.12(71)\\
G 305.36     &    0&    0&  28.45(08)&   4.47(01)& --36.60(01)&   5.78(02)\\
G 308.00     &    0&    0&  13.38(05)&   3.01(01)& --23.22(01)&   3.85(02)&&   0.42(06)&   0.14(02)& --21.63(16)&   2.42(35)\\
G 308.80     &    0&    0&  16.54(07)&   2.73(01)& --33.11(01)&   5.27(03)&&   1.12(09)&   0.15(01)& --32.57(29)&   7.13(62)\\
G 308.92     &    0&    0&  25.96(07)&   5.07(02)& --51.46(01)&   4.55(02)&&   0.18(04)&   0.10(03)& --51.08(18)&   1.66(43)\\
G 309.92     &    0&    0&  22.08(07)&   4.32(02)& --57.60(01)&   4.35(02)\\
G 316.77     &   20&   20&  15.30(08)&   2.51(01)& --41.05(01)&   5.48(04)\\
G 316.81     &    0&   20&  17.51(08)&   2.85(01)& --39.82(01)&   5.63(03)\\
G 318.05     &    0&    0&  27.62(07)&   5.63(02)& --50.48(01)&   4.31(02)\\
G 323.74     &    0&   20&   4.42(06)&   1.11(02)& --50.46(02)&   3.29(06)\\
G 324.20     &    0&   30&  17.16(09)&   2.39(02)& --89.27(02)&   6.11(04)\\
G 326.47     &    0&    0&  12.13(09)&   2.18(02)& --42.61(02)&   4.59(05)&&   1.41(13)&   0.16(02)& --41.39(35)&   8.17(99)\\
G 326.64     &    0&    0&  35.04(08)&   7.97(02)& --40.24(00)&   3.95(01)&&   0.46(08)&   0.10(01)& --39.24(40)&   4.31(71)\\
G 328.24     &    0&    0&  24.60(09)&   3.21(01)& --43.69(01)&   7.11(03)\\
G 328.30     &    0&    0&  25.83(10)&   3.56(02)& --92.92(01)&   6.47(03)\\
G 328.81     &    0&    0&  55.86(15)&   8.60(03)& --42.67(01)&   5.34(02)&&   4.65(10)&   0.60(02)& --41.37(07)&   6.37(20)\\
G 329.03     &    0&    0&  13.89(08)&   2.11(01)& --44.19(02)&   5.57(04)&&   1.71(09)&   0.28(02)& --43.70(15)&   5.69(41)\\
G 330.88     &    0&    0&  43.65(23)&   6.92(04)& --63.38(01)&   5.45(04)&&   2.59(14)&   0.34(02)& --62.78(18)&   7.08(45)\\
G 330.95     &   20&   20&  59.42(16)&   6.79(02)& --92.16(01)&   8.36(03)&&   2.83(12)&   0.27(01)& --91.38(21)&   9.77(44)\\
G 331.28     &   40&  --20&  21.07(09)&   3.86(02)& --89.39(01)&   5.01(03)\\
G 332.83     &    0&  --20&  85.39(13)&  10.90(02)& --57.94(01)&   7.13(01)&&   3.24(13)&   0.44(02)& --57.38(14)&   6.92(33)\\
G 333.13     &    0&    0&  59.60(17)&   8.28(03)& --52.55(01)&   6.13(02)&&   1.07(13)&   0.12(02)& --52.22(38)&   6.89(140)\\
G 333.60     &   20&    0&  40.04(14)&   4.10(02)& --49.60(02)&   8.78(04)\\
G 337.40     &   20&   20&  43.56(10)&   7.48(02)& --41.77(01)&   5.29(01)&&   2.81(16)&   0.37(03)& --40.69(16)&   6.42(49)\\
G 340.06     &    0&    0&  34.59(13)&   4.60(02)& --54.62(01)&   6.59(03)&&   1.25(13)&   0.17(02)& --53.55(37)&   6.97(81)\\
G 343.12     &    0&    0&  16.80(17)&   2.84(03)& --31.08(03)&   5.24(07)&&   0.59(11)&   0.13(02)& --31.12(43)&   4.34(87)\\
G 345.00     &    0&    0&  27.05(40)&   2.74(06)& --27.58(07)&   7.12(16)&&   3.56(13)&   0.39(02)& --26.85(15)&   8.51(38)\\
G 345.01     &    0&    0&  44.61(08)&   6.99(01)& --15.04(01)&   5.79(01)&&   1.39(10)&   0.25(02)& --14.27(18)&   5.28(47)\\
G 345.41     &    0&    0&  36.99(09)&   6.46(02)& --22.11(01)&   5.17(02)\\
G 345.51     &    0&    0&  32.29(13)&   5.03(03)& --18.45(01)&   5.31(03)&&   1.10(12)&   0.22(03)& --17.34(25)&   4.70(60)\\
G 348.55     &    0&    0&  31.34(12)&   4.32(02)& --16.16(01)&   6.36(03)\\
G 348.73     &    0&    0&  76.64(12)&  11.80(02)& --12.68(00)&   5.78(01)&&   1.38(09)&   0.25(02)& --11.16(17)&   5.12(38)\\
G 350.10     &    0&    0&  27.58(11)&   2.61(01)& --68.83(02)&   9.98(04)&&   0.87(14)&   0.07(01)& --68.18(99)&  11.77(160)\\
G 351.41     &    0&    0&  57.97(20)&   9.05(04)&  --7.51(01)&   5.57(03)&&   3.68(20)&   0.42(03)&  --7.60(18)&   7.45(62)\\
G 351.58     &    0&    0&  24.04(12)&   3.34(02)& --96.43(00)&   6.37(04)&&   2.20(10)&   0.36(02)& --94.72(13)&   5.67(31)\\
G 351.78     &    0&    0&  88.86(38)&  10.40(06)&  --3.55(02)&   6.11(04)&&   10.54(15)&   1.08(02)&  --2.95(05)&   8.59(14)\\
G 353.41     &    0&    0&  55.52(10)&   8.52(02)& --16.35(01)&   6.05(01)&&   1.30(12)&   0.25(02)& --16.11(22)&   4.92(52)\\
G 359.97     &    0&    0&  20.16(09)&   5.28(02)&  17.71(01)&   3.57(02)\\
Orion KL   &    0&    0&  53.52(135)&   5.16(23)&   7.69(11)&   6.99(35)&&  52.50(70)&   3.90(08)&   7.16(07)&  10.38(23)\\
Sgr A        &    0&    0&  31.22(48)&   1.50(03)&  11.96(15)&  17.51(37)&&  12.95(25)&   1.07(02)&  14.19(10)&  10.96(26)\\
S 231        &    0&    0&   8.10(09)&   1.55(02)& --17.10(03)&   4.71(07)&&   0.36(08)&   0.15(03)& --15.45(26)&   2.25(55)\\
S 255        &    0&    0&  15.50(06)&   3.50(02)&   6.67(01)&   3.97(02)&&   0.53(07)&   0.15(02)&   8.24(23)&   3.26(47)\\
\noalign{\smallskip}
\hline
\end{tabular}
\normalsize
\end{flushleft}
\label{table:1.4mm}
\end{table*}

\begin{table*}
\caption[]{C$^{18}$O ($1-0$) and HNCO ($5_{05}-4_{04}$) integrated intensities 
and gaussian line parameters at the indicated positions (cf. 
Tables~\ref{table:sest},\ref{table:oso})
measured at Onsala.
The numbers in the brackets are the statistical
uncertainties in the last digits (standard deviations).}
\begin{flushleft}
\begin{tabular}{lrrlrrlllrrl}\hline\noalign{\smallskip}
&&&\multicolumn{4}{l}{C$^{18}$O ($1-0$)}
&&\multicolumn{4}{l}{HNCO ($5_{05}-4_{04}$)}
\\
\cline{4-7}
\cline{9-12}
\noalign{\smallskip}
Source
&$\Delta \alpha$  &$\Delta \delta$
&$\int T_{\rm mb} dv$
&\multicolumn{1}{l}{$T_{\rm mb}$}
&\multicolumn{1}{l}{$V_{\rm LSR}$}
&$\Delta V$
&
&$\int T_{\rm mb} dv$
&\multicolumn{1}{l}{$T_{\rm mb}$}
&\multicolumn{1}{l}{$V_{\rm LSR}$}
&$\Delta V$
\\
&(\arcsec) &(\arcsec)
&(K$\cdot$km/s)
&\multicolumn{1}{l}{(K)} 
&\multicolumn{1}{l}{(km/s)}
&\multicolumn{1}{l}{(km/s)}
&
&(K$\cdot$km/s)
&\multicolumn{1}{l}{(K)} 
&\multicolumn{1}{l}{(km/s)}
&\multicolumn{1}{l}{(km/s)}
\\
\noalign{\smallskip}
\hline\noalign{\smallskip}
G 121.30     &    0&    0&   5.94(18)&   1.30(07)& --17.32(09)&   3.55(22)&&   0.98(14)&   0.40(06)& --17.57(16)&   2.30(37)\\
G 123.07     &    0&    0&   3.19(14)&   0.78(06)& --30.28(08)&   3.81(32)&&   1.07(20)&   0.12(02)& --31.14(84)&   8.21(182)\\
G 133.69     &    0&  --40&   8.39(20)&   1.22(03)& --42.38(08)&   5.94(19)&&   0.55(14)&   0.14(03)& --43.37(51)&   3.77(106)\\
G 133.95     &    0&    0&  10.07(19)&   1.71(04)& --47.60(05)&   5.12(13)&&   1.38(24)&   0.13(03)& --46.75(77)&   9.95(231)\\
G 170.66     &    0&    0&   2.19(20)&   0.56(05)& --15.27(24)&   3.37(35)\\
G 173.17     &    0&    0&   4.50(27)&   0.87(07)& --19.83(13)&   4.78(40)\\
G 173.48     &    0&    0&   3.85(19)&   0.76(04)& --16.22(12)&   4.77(28)&&   1.25(20)&   0.33(08)& --15.98(20)&   2.68(69)\\
G 173.72     &    0&    0&   3.32(20)&   0.97(09)& --16.83(12)&   3.64(38)&&   0.49(14)&   0.21(06)& --17.58(32)&   2.22(72)\\
G 188.95     &    0&    0&   3.40(33)&   0.70(09)&   2.93(20)&   4.54(66)\\
G 34.26      &    0&    0&  31.00(28)&   4.45(04)&  57.80(03)&   6.47(07)&&    2.89(20)&   0.54(04)&   30.73(18)&   5.55(41)\\
G 40.50      &    0&    0&   4.69(13)&   0.81(03)&  32.61(07)&   5.22(20)&&   0.60(20)&   0.14(08)&  32.54(47)&   3.94(237)\\
G 43.17      &    0&    0&  25.71(40)&   1.45(02)&   7.91(13)&  14.84(28)&&   3.31(28)&   0.24(02)&   8.77(56)&  12.97(118)\\
G 49.49      &    0&    0&  48.02(50)&   2.99(04)&  57.82(08)&  12.45(19)&&   7.93(30)&   0.69(03)&  56.77(19)&  10.20(50)\\
G 60.89      &    0&    0&   6.17(24)&   1.33(07)&  22.40(07)&   4.35(26)&&   0.91(25)&   0.14(03)&  21.15(94)&   7.07(143)\\
G 61.48      &    0&    0&   4.65(19)&   1.12(06)&  21.85(09)&   3.82(23)\\
G 69.54      &    0&    0&   9.26(21)&   1.98(05)&  11.19(04)&   4.52(14)&&   2.80(22)&   0.43(04)&  11.04(20)&   5.38(57)\\
G 70.29      &    0&    0&   6.06(16)&   0.70(02)& --24.28(10)&   7.60(25)\\
G 75.78      &    0&    0&   8.96(18)&   1.43(04)&  --0.04(06)&   5.28(14)&&   1.58(22)&   0.16(02)&  --1.88(66)&   9.28(145)\\
G 77.47      &    0&    0&   5.59(16)&   1.16(03)&   0.50(07)&   3.98(13)&&   1.00(20)&   0.14(03)&   0.98(72)&   6.75(151)\\
G 81.72      &    0&    0&  19.07(22)&   3.74(04)&  --2.77(03)&   4.49(06)&&   3.20(20)&   0.53(04)&  --2.71(19)&   5.55(41)\\
G 81.77      &    0&    0&   8.23(13)&   1.97(05)&  --4.16(04)&   3.65(11)&&   1.12(12)&   0.25(04)&  --3.97(19)&   3.71(57)\\
G 81.87      &    0&    0&  13.45(24)&   2.44(05)&   9.36(05)&   4.73(10)&&   2.37(28)&   0.36(06)&   9.52(31)&   5.61(106)\\
G 92.67      &    0&    0&   4.22(16)&   1.12(03)&  --6.12(08)&   3.66(12)&&   0.82(14)&   0.21(04)&  --5.26(37)&   3.68(70)\\
G 99.98      &    0&    0&   5.21(14)&   1.33(04)&   0.67(06)&   3.64(11)&&   0.46(10)&   0.23(07)&   0.67(15)&   1.21(39)\\
G 108.76--0.95&    0&    0&   6.79(16)&   1.69(08)& --50.78(04)&   3.77(20)&&   0.63(12)&   0.23(06)& --50.63(22)&   2.00(54)\\
G 108.76--0.98&    0&    0&  11.97(17)&   3.09(05)& --51.13(04)&   3.51(06)&&   0.67(16)&   0.16(04)& --51.08(45)&   4.01(110)\\
G 111.53     &    0&    0&  12.22(22)&   1.80(04)& --56.23(05)&   6.18(14)&&   2.74(20)&   0.55(05)& --56.44(16)&   4.66(42)\\
\noalign{\smallskip}
\hline
\end{tabular}
\normalsize
\end{flushleft}
\label{table:oso-rst}
\end{table*}

\begin{figure}
\centering
\resizebox{\hsize}{!}{\includegraphics{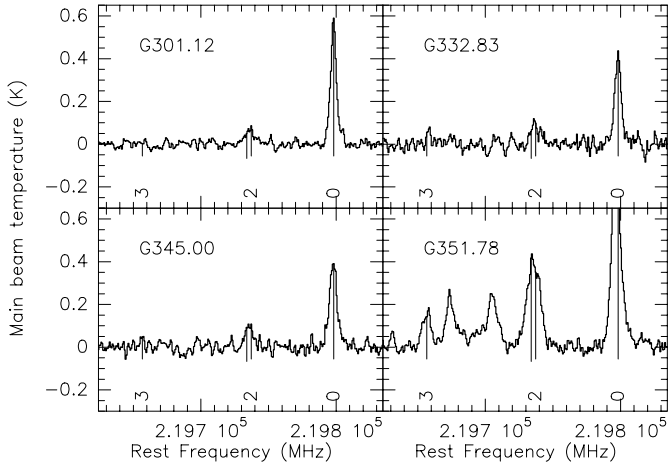}}
\caption{Examples of 220 GHz HNCO spectra obtained at SEST 
covering  $K_{-1}=0$, 2 and 3 transitions}
\label{fig:examples}
\end{figure}

\begin{figure*}
\begin{minipage}[b]{0.45\textwidth}
\centering
\resizebox{\hsize}{!}{\includegraphics{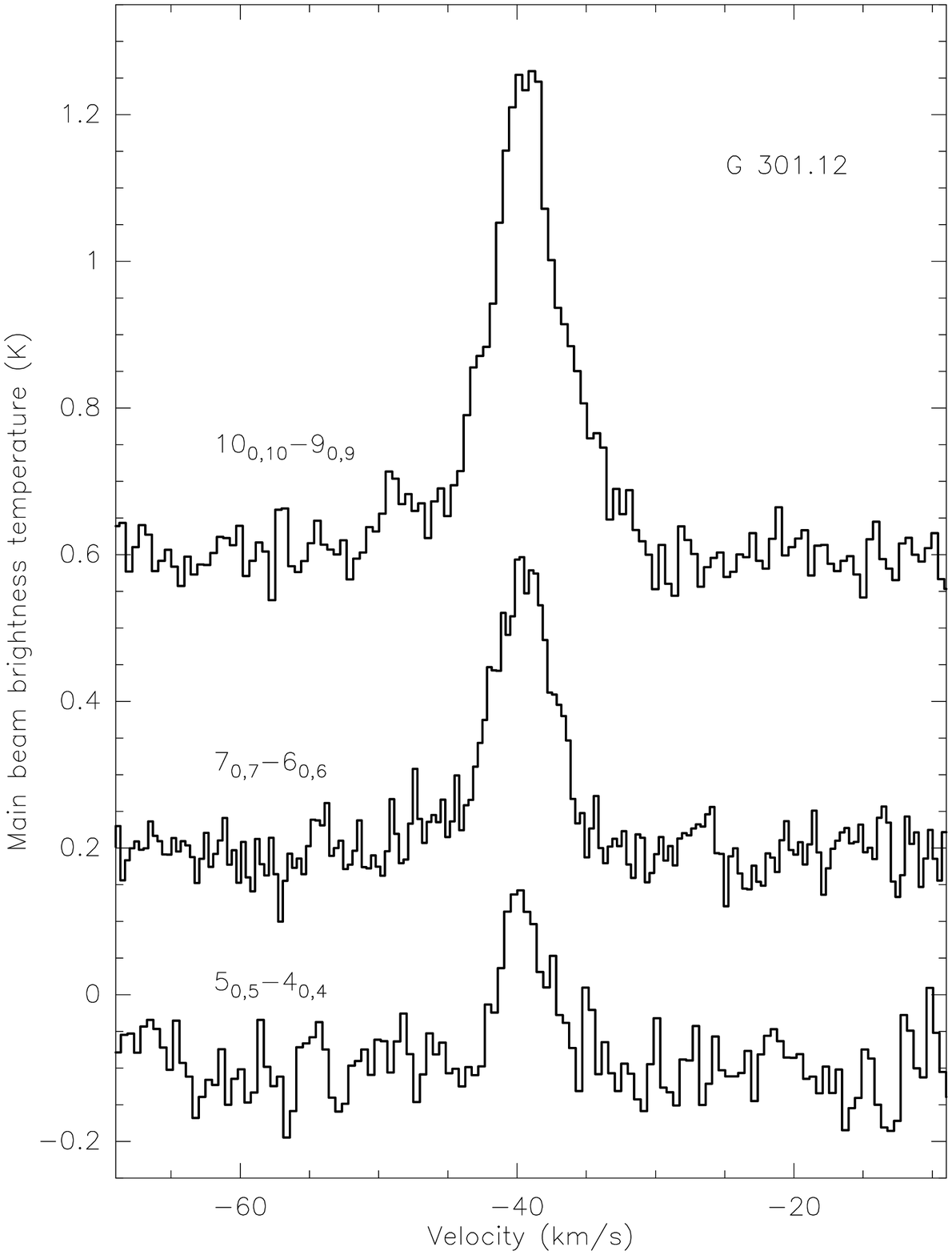}}
\end{minipage}
\hfill
\begin{minipage}[b]{0.45\textwidth}
\centering
\resizebox{\hsize}{!}{\includegraphics{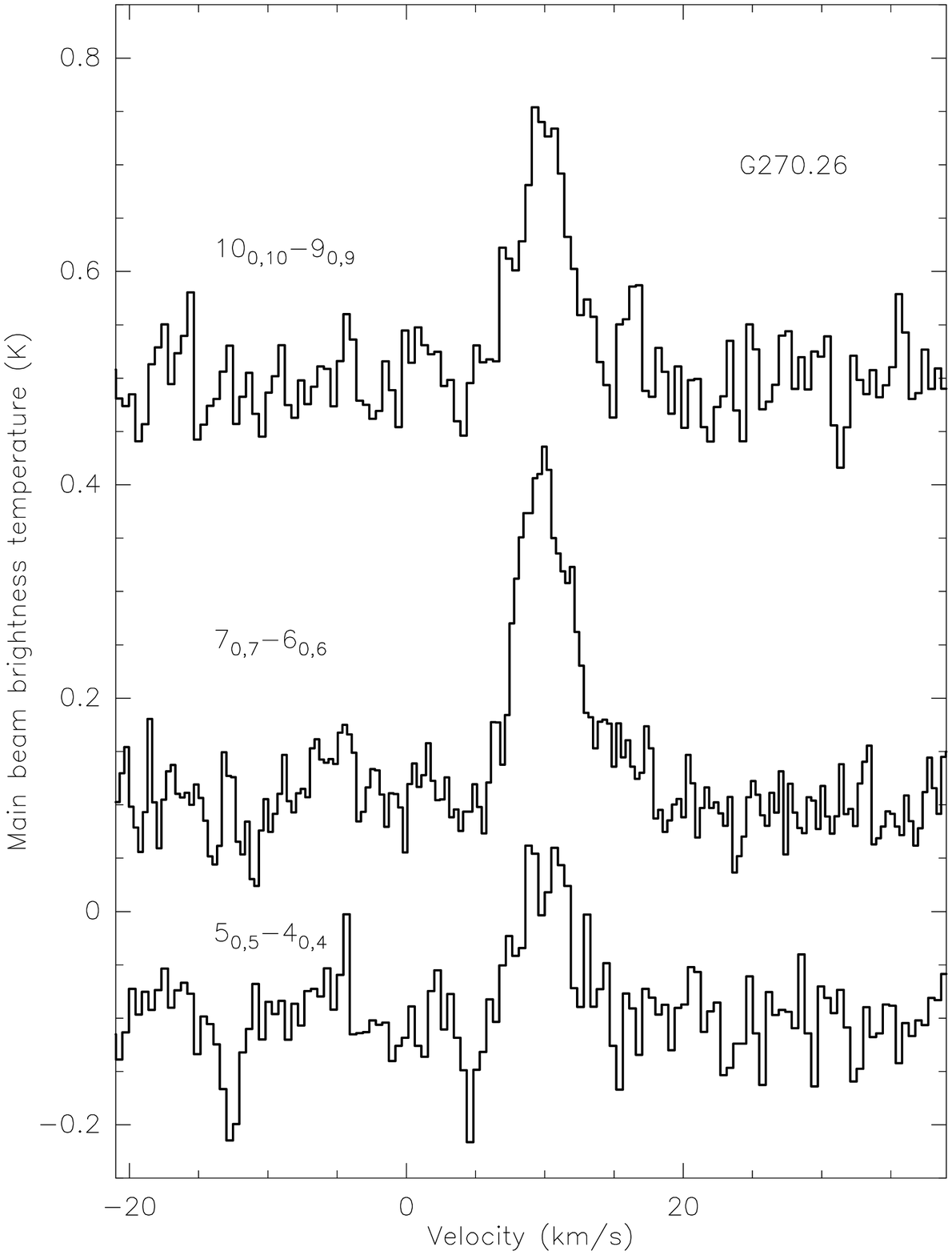}}
\end{minipage}\\[7mm]
\begin{minipage}[b]{0.45\textwidth}
\centering
\resizebox{\hsize}{!}{\includegraphics{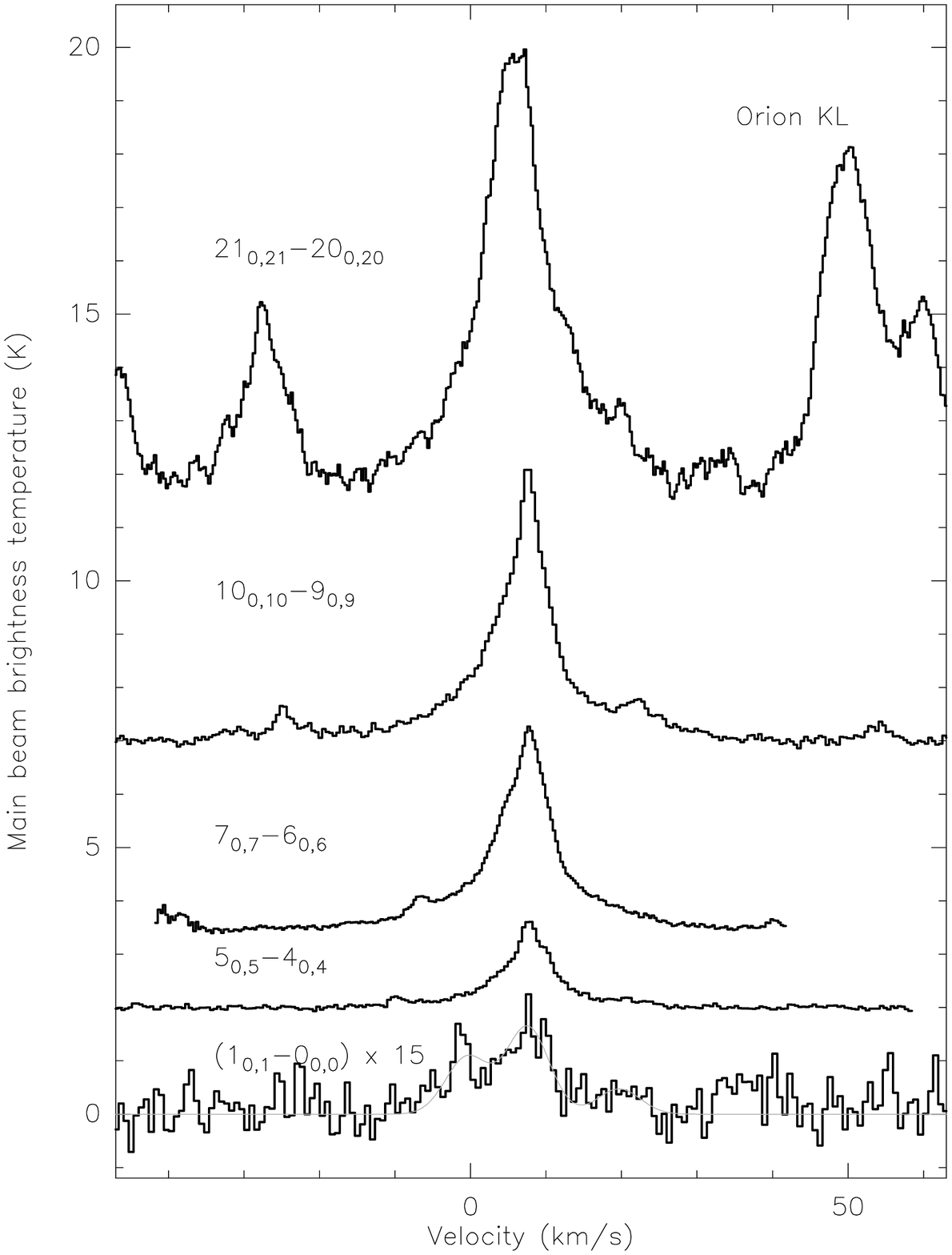}}
\end{minipage}
\hfill
\begin{minipage}[b]{0.45\textwidth}
\centering
\resizebox{\hsize}{!}{\includegraphics{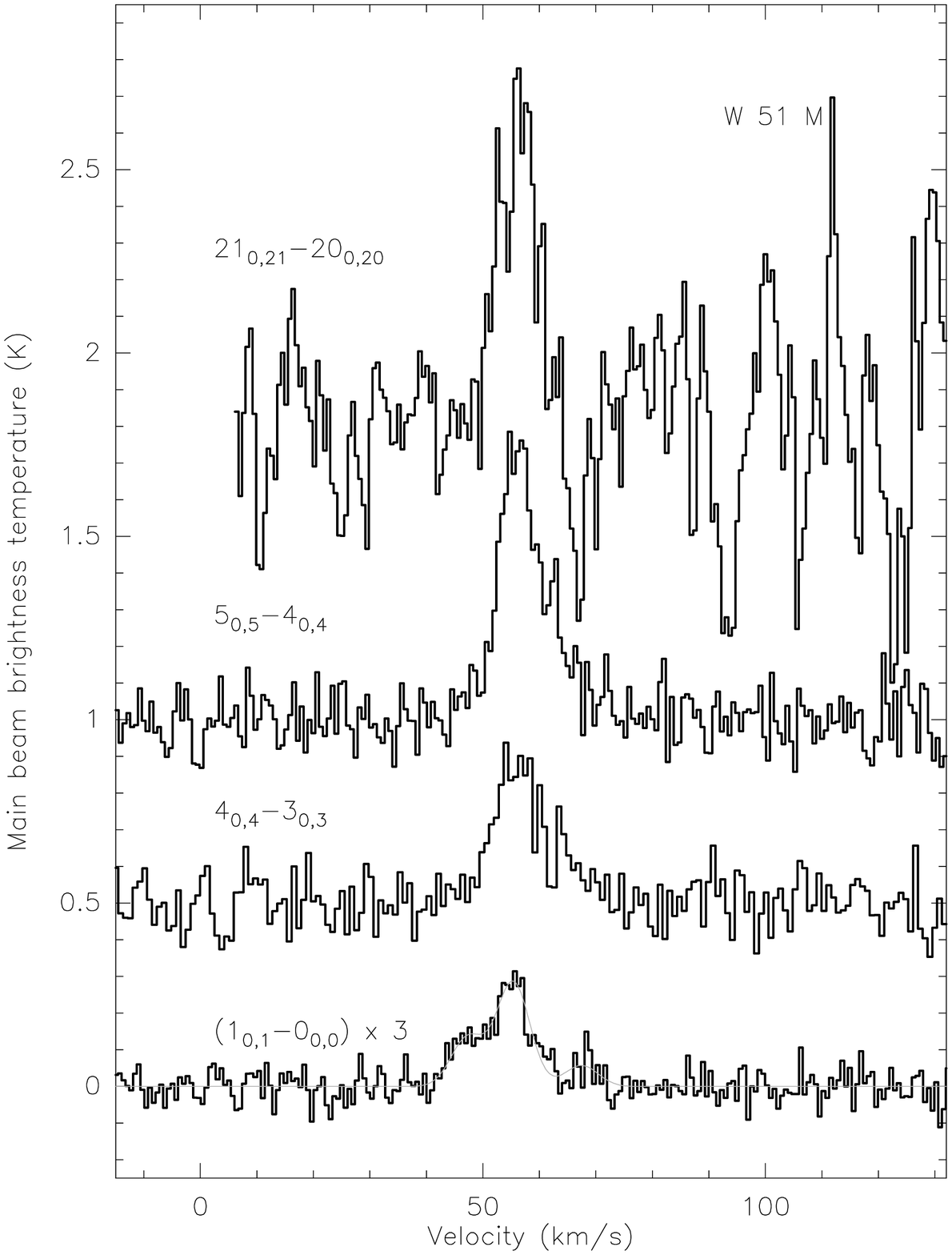}}
\end{minipage}
\caption{HNCO $K_{-1}=0$ lines in four selected sources. 
For the $1_{01}-0_{00}$ transition a 3-component gaussian 
(according for hyperfine structure) is superposed}
\label{fig:over}
\end{figure*}

\addtocounter{table}{8}
Examples of measured spectra are given in 
Figs.~\ref{fig:examples}, \ref{fig:over}. Fig.~\ref{fig:examples} 
shows spectra of a few
sources covering $K_{-1}=0$, 2 and 3 transitions at 220~GHz. 
Fig.~\ref{fig:over} presents HNCO spectra
in the HNCO $K_{-1}=0$ transitions at different wavelengths for
several sources.

The HNCO line profile in Orion~KL can be decomposed into at least
two components which likely correspond to the so-called classical
``Hot Core'' and ``Plateau'' outflow components (see, e.g., Harris
et al. 1995). The ratio between these components is practically the same
for the $5_{05}-4_{04}$, $7_{07}-6_{06}$ and $10_{0,10}-9_{0,9}$ lines:
$\sim 60$\% of the emission originates from the ``Plateau'' outflow source.
The other lines do not allow such decomposition due to their weakness
or blending with other spectral features.

An inspection of 
Table~\ref{table:1.4mm} shows that the derived C$^{18}$O velocities are
systematically lower (more negative) than the HNCO ones. The difference
is $\sim -1$~km/s on the average. This can be an instrumental effect:
the C$^{18}$O line was located far away from the center of the 
spectrometer band and a possible non-linearity in the frequency response
could lead to the apparent displacement of the line on the velocity 
axis. This remark is applicable also for the higher $K_{-1}$  HNCO lines.

\subsection{Maps}

\begin{figure*}
\begin{minipage}[b]{0.32\textwidth}
\centering
\resizebox{\hsize}{!}{\includegraphics{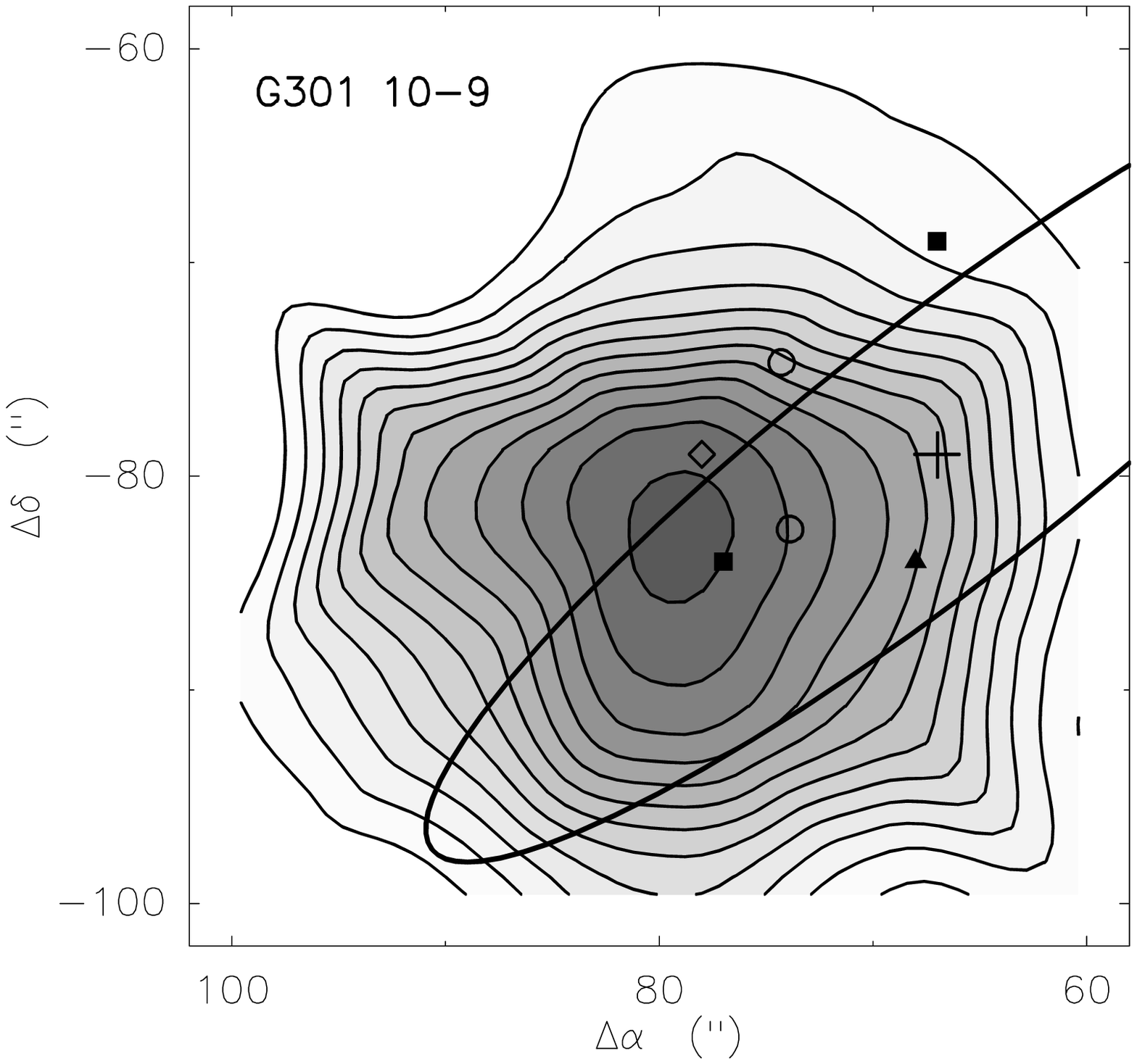}}
\end{minipage}
\hfill
\begin{minipage}[b]{0.32\textwidth}
\centering
\resizebox{\hsize}{!}{\includegraphics{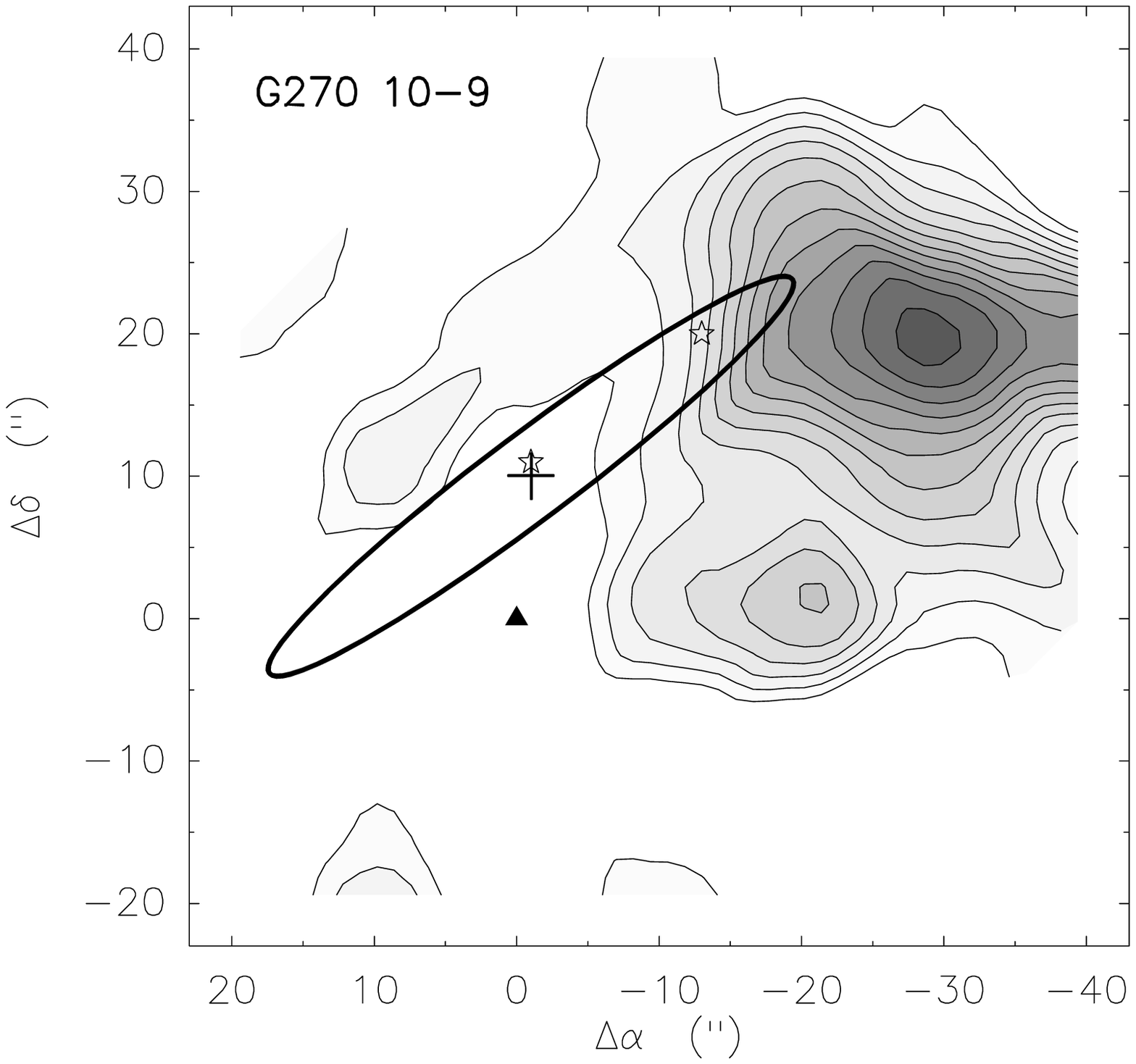}}
\end{minipage}
\hfill
\begin{minipage}[b]{0.32\textwidth}
\centering
\resizebox{\hsize}{!}{\includegraphics{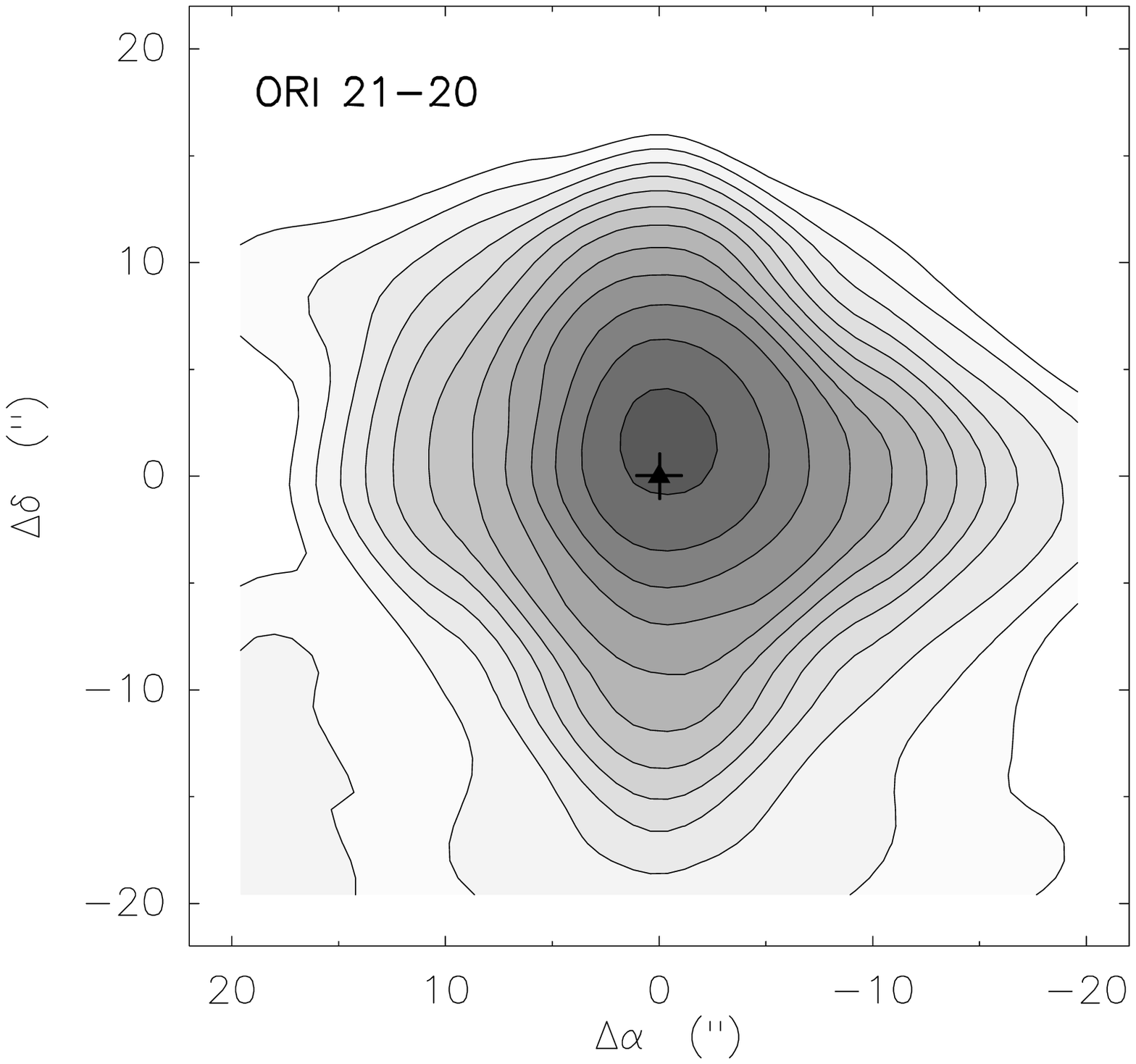}}
\end{minipage}
\caption{The HNCO $10_{0,10}-9_{0,9}$ integrated intensity maps of
G301.12--0.20 and G270.26+0.83 and HNCO $21_{0,21}-20_{0,20}$ 
integrated intensity map of Orion KL.
The levels start from 15\% of the peak intensities in steps of 7.5\%.
The peak intensities equal to 4.4, 1.5 and 95.0 K$\cdot$km\,s$^{-1}$ for
G301.12--0.20, G270.26+0.83 and Orion KL, respectively. The beam widths
are 24\arcsec\ for the first two objects and 18\arcsec\ for Orion KL.
For the first two objects
the large crosses indicate IRAS positions, small stars show NIR sources,
triangles mark H$_2$O masers, squares correspond to OH masers and
diamonds show methanol masers (for references see Lapinov et al. 1998).
Open circles mark UC H\,{\sc ii} regions (Walsh et al. 1998).
The IRAS uncertainty ellipses are shown.
For Orion KL only the IRc~2 position is indicated}
\label{fig:maps}
\end{figure*}

In order to estimate source sizes and their spatial association with YSO
and infrared (IR) sources we mapped 2 southern sources
in the $10_{0,10}-9_{0,9}$ HNCO line and 
Orion KL, W49N and W51M
in the $15_{0,15}-14_{0,14}$, $16_{0,16}-15_{0,15}$ and 
$21_{0,21}-20_{0,20}$ lines. 
W51M was mapped also in the
$5_{05}-4_{04}$ line. Three of these maps are presented in 
Fig.~\ref{fig:maps}.

The sources remain spatially unresolved. E.g. for G~301.12--0.20
we obtain a FWHM $\approx 29$\arcsec\ in right ascension (from the strip scan
across the map) which is very close
to the beam size at this frequency (24\arcsec). 

\subsection{Detection of the $K_{-1}=5$ HNCO transition}
\label{sec:k5}

\begin{figure*}
\begin{minipage}[b]{0.48\textwidth}
\centering
\resizebox{\hsize}{!}{\includegraphics{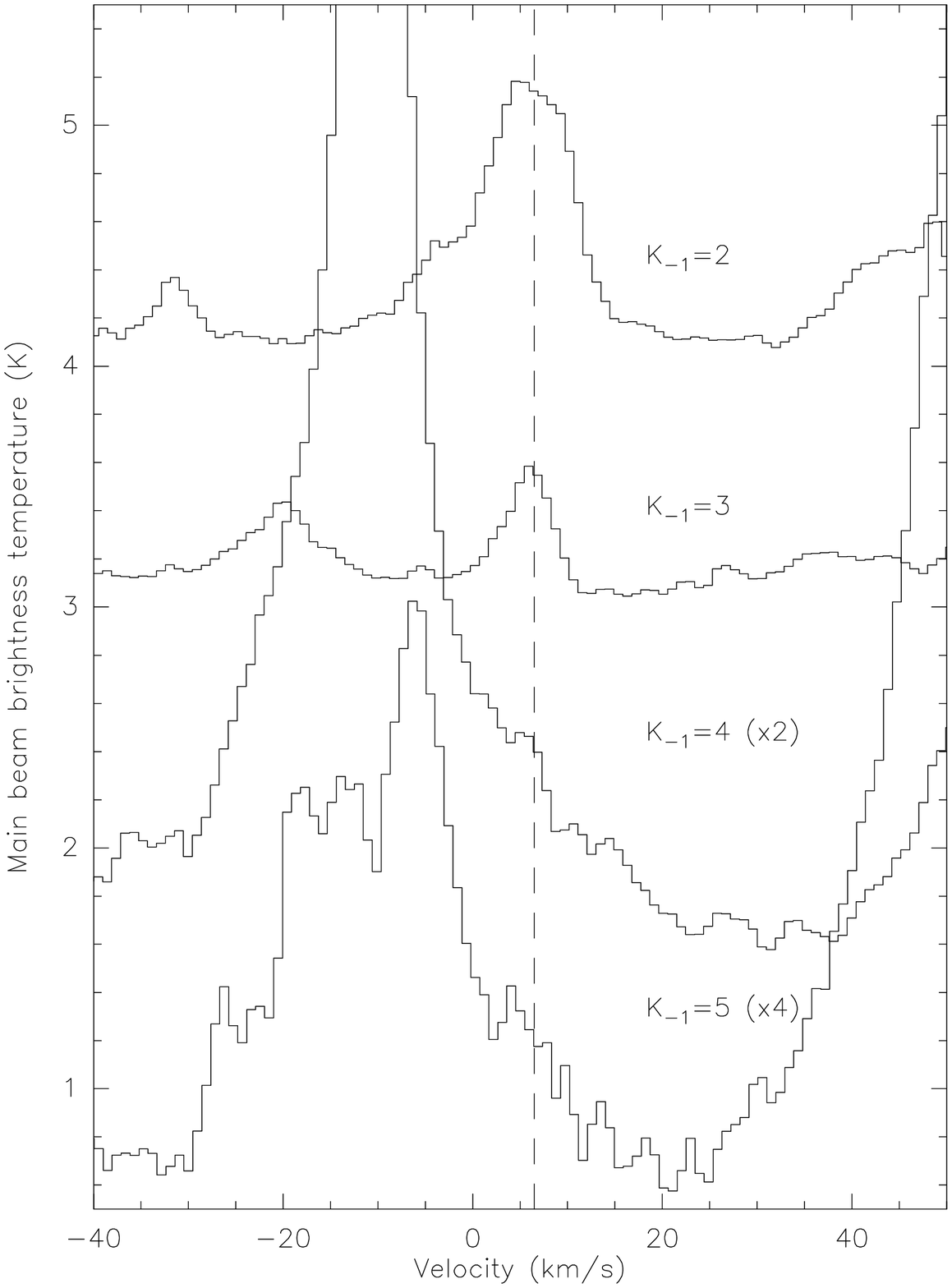}}
\end{minipage}
\hfill
\begin{minipage}[b]{0.48\textwidth}
\centering
\resizebox{\hsize}{!}{\includegraphics{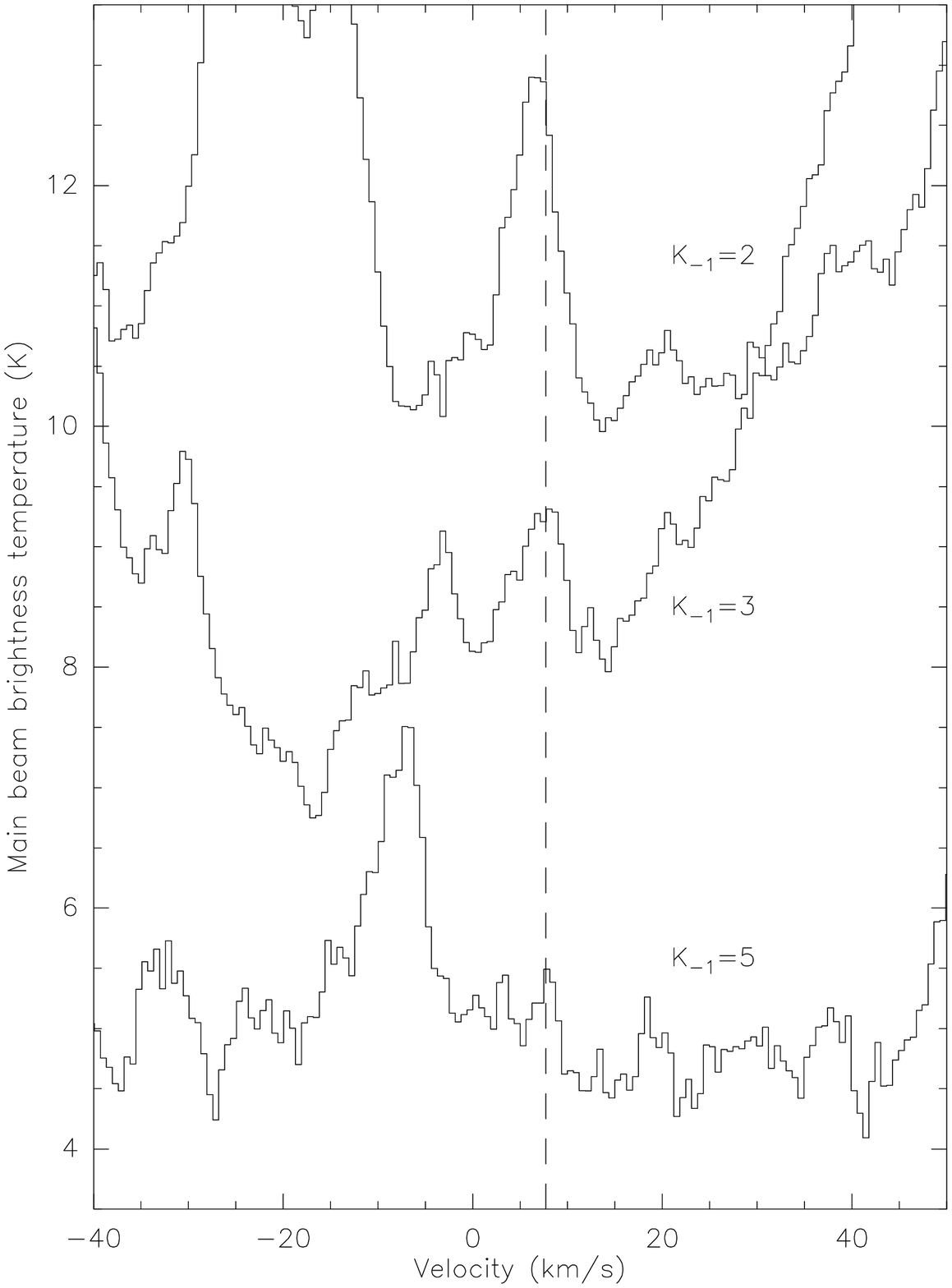}}
\end{minipage}
\caption{Left panel: 
parts of the Orion low-resolution 220~GHz spectrum corresponding
to higher $K_{-1}$ HNCO transitions. The profiles are aligned in
velocity. No baselines are subtracted but the subspectra are shifted
along the $y$-axis for clarity. The dashed vertical line corresponds
to $V_{\rm LSR}=6.5$~km\,s$^{-1}$. Right panel: the same for the 461~GHz
spectrum. Here the dashed vertical line corresponds
to $V_{\rm LSR}=7.5$~km\,s$^{-1}$}
\label{fig:ori-mk}
\end{figure*}

The highest $K_{-1}$ HNCO transition reported so far was $K_{-1}=4$
(the $10_4-9_4$ line) in Orion (Sutton et al. 1985). This line is located
on the shoulder of the strong C$^{18}$O $J=2-1$ line. In Fig.~\ref{fig:ori-mk}
we show parts of our Orion 220~GHz low resolution spectrum and 461~GHz
spectra with $K_{-1}=2$, 3, 4 and even 5 features
(the $K_{-1}=1$ transition is outside our band).
The rest frequencies are assumed to be equal to those given in the 
JPL catalogue for the strongest components of the corresponding transitions
(for $K_{-1}=2$ at 220~GHz we took the mean of the frequencies of the 
two strongest components).

There is a weak bump in the redshifted C$^{18}$O $J=2-1$ 
wing which can be attributed to HNCO $10_4-9_4$.
Due to the uncertainty in fitting the C$^{18}$O line profile the intensity
of the HNCO feature cannot be reliably determined but it is lower than
reported by Sutton et al. (1985). Our best estimate for the integrated
intensity is $\int T_{\rm mb}dv\sim 0.7$~K\,km/s, but a reliable error
cannot be given.

There is also a feature at the $K_{-1}=5$ frequency in the 220~GHz spectrum.
It is located in the wing of a C$_2$H$_3$CN line. The integrated
intensity is $\int T_{\rm mb}dv = 0.27\pm 0.06$~K\,km/s. The identification
of this feature with HNCO seems to be reliable. The only other candidate is
the C$_2$H$_5$OH $14_{21,2}-13_{11,2}$ line at 219391.81~MHz.
However, there is no sign of other ethanol lines in
our spectrum so we reject this alternative.
In the 461~GHz spectrum the $K_{-1}=5$ feature is clearly detected.
Its integrated
intensity is $\int T_{\rm mb}dv = 1.4\pm 0.2$~K\,km/s.

\subsection{Hyperfine splitting, HN$^{13}$CO and optical depths}

The HNCO lines are split into several hyperfine components mainly due
to the $^{14}$N spin. This splitting is clearly seen in the
$1_{01}-0_{00}$ transition (Fig.~\ref{fig:over}) at 22~GHz. Earlier
HNCO hyperfine structure in the $1_{01}-0_{00}$ line
was only observed in the dark cloud TMC-1 (\cite{brown81}) where
possible deviations from the optically thin LTE (Local Thermodynamic 
Equilibrium) intensity ratios (3:5:1)
were found. In our spectra the hyperfine ratios are consistent with the 
optically thin LTE values. 
Taking into account the measurement uncertainties, an upper limit
on the optical depth in this transition for the sources detected in 
Effelsberg is $\tau\la 1$.

\begin{figure}
\resizebox{\hsize}{!}{\includegraphics{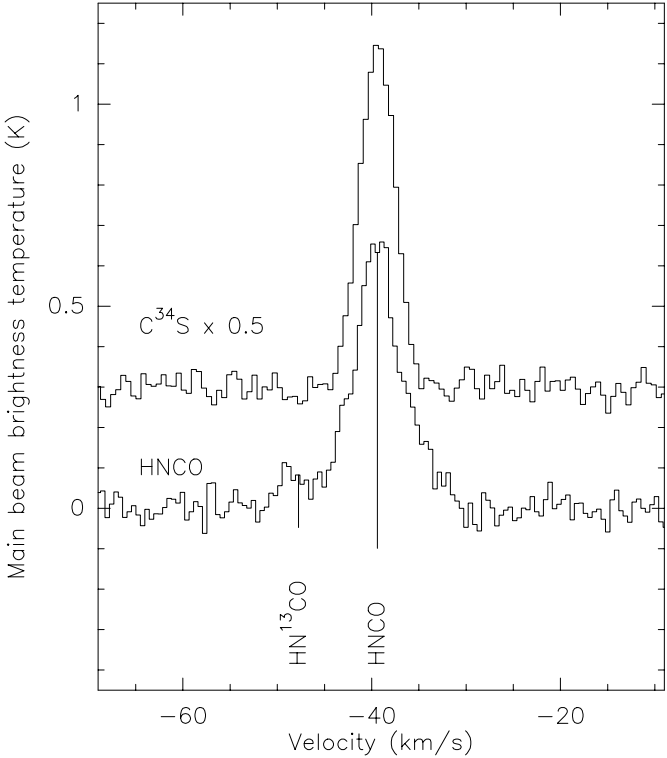}}
\caption{The HNCO $10_{0,10}-9_{0,9}$ line in 
G301.12--0.20 in comparison 
with the C$^{34}$S(2--1) line. The expected location of the HN$^{13}$CO 
$10_{0,10}-9_{0,9}$ line is shown}
\label{fig:g301-hn13co}
\end{figure}

To the best of our knowledge no isotopomer of HNCO except the main one
has been detected in space yet. This detection would be important 
for estimates of HNCO optical depths which are believed
to be small (e.g. Jackson et al. 1984, Churchwell et al. 1986).
The frequency separations between the HN$^{13}$CO and the main isotopomer
lines are rather small corresponding to a few km/s, so in sources with
broad lines like Orion~A or Sgr~A the HN$^{13}$CO lines will be blended.
However, there are some strong HNCO sources in our sample with
narrower lines which show features attributable to HN$^{13}$CO. The most
reliable one is seen in the G~301.12--0.20 $10_{0,10}-9_{0,9}$ spectrum
(Fig.~\ref{fig:g301-hn13co}). A weak feature on the blue shoulder of the
main isotope line is very close in frequency to the expected location of
the HN$^{13}$CO line .

For comparison we show in addition to 
HNCO also the C$^{34}$S spectrum. It is noteworthy that there is no
bump in this spectrum corresponding to the discussed feature in HNCO.

The line we identify with HN$^{13}$CO
is shifted by $0.65\pm 0.21$~MHz from the 
expected HN$^{13}$CO transition frequency. 
This $3\sigma$ shift, if it is significant, cannot be explained 
by instrumental effects like in the case of our C$^{18}$O data
because the feature is very close to the main isotope line.
The shift greatly exceeds the uncertainty of the transition frequency
derived from the laboratory data (Winnewisser et al. 1978) which is 25~kHz.
This makes the identification questionable. Detection of other
HN$^{13}$CO lines would be important in this respect.
There is no corresponding feature in the $7_{07}-6_{06}$ HNCO spectrum
(the $5_{05}-4_{04}$ spectrum is too noisy). This could mean that the
optical depth in this transition is significantly lower. Indeed, at
sufficiently high temperatures ($> 30$~K) it can be about 2 times lower
than in the $10_{0,10}-9_{0,9}$ transition according to Eq.~(\ref{eq:tau-r}) 
(see the discussion in Sect.~\ref{sec:rd}).

If our identification of the discussed line with HN$^{13}$CO is correct
we can estimate the optical depth assuming the same excitation as for 
the main isotopomer. For G~301.12--0.20 we obtain $\tau(\mbox{HNCO})
\approx 15$ if we assume the terrestrial $^{12}$C/$^{13}$C isotope 
ratio ($^{12}$C/$^{13}$C = 89) and $\approx 7$ for $^{12}$C/$^{13}$C = 40. 
A high optical depth in the $10_{0,10}-9_{0,9}$ HNCO line does not contradict
our conclusion of low optical depth in the $1_{01}-0_{00}$ transition because
the line strengths for these transitions are different (see discussion
in Sect.~\ref{sec:rd}).
Therefore, the optical
depth in {\it some} lines of the main isotopomer might be rather high. This
contradicts the usual assumption of low optical depth in all HNCO lines
(e.g., Jackson et al. 1984, Churchwell et al. 1986) and could imply serious
consequences for the analysis of HNCO excitation and abundances.

\section{Discussion} \label{sec:general}

\subsection{Comparison with other HNCO data}\label{sec:comp}

Most of our HNCO sources are new detections. Only few 
were included in the surveys of Jackson et al. (1984) and
Churchwell et al. (1986). 
A direct comparison with the intensities measured
by Jackson et al. is impossible due to different temperature scales.
Common detected sources are Orion~KL and W51. 
Their upper limit for
W3(OH) does not contradict our value if we take into account the 
difference in the temperature scales. 
The upper limits for the $1_{01}-0_{00}$ transition obtained by Churchwell
et al. do not contradict our results taking into account the differences
in the beam sizes and efficiencies.

As mentioned above, towards Orion KL several HNCO lines were observed 
at 220~GHz by Sutton et al. (1985). 
Their results agree in general with our measurements
though there is a discrepancy concerning the intensity of the $K_{-1}=4$
transition (Sect.~\ref{sec:k5}).

It is worth noting that while at 22~GHz
and at 110~GHz (as obtained by Jackson et al. 1984) the brightest source of
HNCO emission is the Galactic center, at 220~GHz 
the situation changes
and Orion becomes the brightest source with several other sources
approaching Sgr~A in intensity. Apparently this is caused by differences
in excitation. 

\subsection{Rotational diagrams}\label{sec:rd}

As a first step in the excitation analysis we construct traditional
rotational diagrams for our sources. For a recent discussion of this
method see e.g. Goldsmith \& Langer (1999). This means a plot of the
column density ($N_{\rm u}$) per statistical weight ($g_{\rm u}$) 
of a number of molecular energy
levels, as a function of their energy above the ground state ($E_{\rm
u}$).  In local
thermodynamic equilibrium (LTE), this will just be a Boltzmann
distribution, so a plot of $\ln(N_{\rm u}/g_{\rm u})$ versus $E_{\rm
u}/k$ will yield a straight line with a slope of $1/T_{\rm R}$. The
temperature inferred is often called the ``rotational temperature''.

Actually from the measurements we do not obtain directly the column
densities. The measured quantity is the line intensity. In an optically
thin case 
for $T_{\rm ex} >> T_{\rm bg}$ ($T_{\rm ex}$ is the excitation 
temperature of
the transition and $T_{\rm bg}$ is the background temperature) 

\begin{equation}
\log\left[ \frac{3k(W/f_{\rm b})}{8\pi^3\nu\mu_{\rm x}^2S}\right] = 
\log\left(\frac{N}{Q}\right) - \frac{E_{\rm u}}{kT_{\rm R}}\log e
\label{eq:rd}
\end{equation}
where $W$ is the integrated line intensity,
$f_{\rm b}$ is the beam dilution factor, $S$ is the line strength,
$\mu_{\rm x}$ is the appropriate component of the dipole moment,
$N$ is the total column density and $Q(T_{\rm R})$ is the
partition function. 

The quantity on the left hand side of Eq.~(\ref{eq:rd}) can be derived
from the molecular data. Plotting it versus
$E_{\rm u}$ we can find the rotational temperature (from the slope)
and the total column density (from the intercept).

Some problems can arise from an uncertainty in the beam filling
factor. As shown in Fig.~\ref{fig:maps} the sources are probably unresolved. 
Assuming that the source size is the same for all HNCO transitions
in a given source and that the source size is small with respect to
the beam, we reduced all data to the same beam size, the SEST HPBW
at 220~GHz, i.e. 24\arcsec.

\begin{figure}
\resizebox{\hsize}{!}{\includegraphics{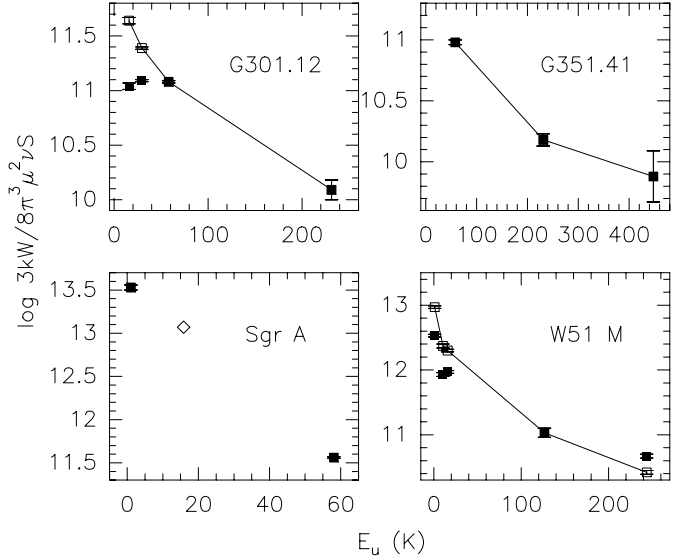}}
\caption[]{Rotational diagrams for selected sample sources
($W=\int T_{\rm mb}dv$, S is the line strength). Filled squares 
correspond to the measured values and the open squares to the values
corrected taking into account the beam sizes (see text). The diamond
on the Sgr~A plot corresponds to the data from Lindqvist et al. 1995}
\label{fig:rd-ex}
\end{figure}

\begin{figure}
\resizebox{\hsize}{!}{\rotatebox{-90}{\includegraphics{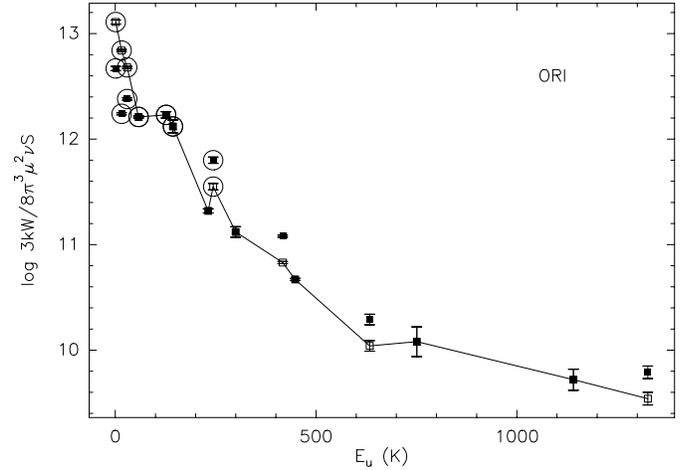}}}
\caption[]{Logarithm of integrated line intensity divided by the line
strength and frequency versus upper-state excitation energy for Orion A.
The filled squares correspond to the measured values and the open squares
represent the values corrected for beam width ratios (see text). 
The points corresponding to $K_{-1}=0$ transitions are encircled
}
\label{fig:ori-rd}
\end{figure}

For Orion the highest observed transition lies $\sim 1300$~K 
above the ground level. For other
sources we managed to observe transitions up to $\sim 450$~K above the 
ground state. Examples of the 
rotational diagrams are presented in Figs.~\ref{fig:rd-ex}, \ref{fig:ori-rd}.

The measured integrated intensities are represented by filled
squares ($f_{\rm b}=1$). The corrected
results are plotted by open squares in Figs.~\ref{fig:rd-ex}, 
\ref{fig:ori-rd}.
One can see that they much better correspond to each other than the
uncorrected values.

The rotational diagram for Orion is presented in
Fig.~\ref{fig:ori-rd}. 
The rotational temperature from this plot is $T_{\rm rot}\approx 25$~K
for the lowest transitions and $T_{\rm rot}\approx 530$~K for the highest
transitions. The latter one is a very high value even for Orion~KL.
But in principle the diagram shows a range of rotational temperatures. We
represent it by 3 components as shown in Table~\ref{table:rot}. 
A separate fit to the $K_{-1}=0$ transitions gives 
$T_{\rm rot}\approx 80$~K (although this fit is not very satisfactory).

The rotational temperatures and column densities derived from 
rotational diagrams are summarized in Table~\ref{table:rot}. 
In this analysis we assume that the sources are optically thin in the
observed transitions. This contradicts the tentative detection of
HN$^{13}$CO in G~301.12--0.20. The effects of high optical depth on
rotational diagrams have been analyzed recently by Goldsmith \& Langer
(1999). 
In optically thick case the column density in the upper level of
the transition ($N_{\rm u}$) is underestimated by the factor of
$\tau/(1-e^{-\tau})$ and, therefore, corresponding points in the
population diagram lie lower than they should.
In general, for linear molecules it produces a curvature resembling 
that seen in
the diagrams for Orion and some other sources. It is caused by the fact
that the optical depth exhibits a peak for transitions with the excitation
energy $E_{\rm u}\sim kT$ (Goldsmith \& Langer 1999). However, for 
nonlinear molecules the optical depth effect rather leads to a ``scatter'' 
in the population diagram, because transitions with significantly
different optical depth can have similar excitation energies.

There is a
strong argument {\it against} high optical depth at least for transitions
with $E_{\rm u}\sim 200-400$~K in Orion. In this range
transitions with 
similar energies of the upper state but with very different frequencies
(belonging to different $K_{-1}$ ladders) were observed. It is easy
to estimate the expected ratio of peak optical depths in the lines
which is 
\begin{equation}
\frac{\tau_1}{\tau_2}=\frac{S_1}{S_2}\,
\frac{\exp\left( {\frac{h\nu_1}{kT}}\right) -1}
{\exp\left( {\frac{h\nu_2}{kT}}\right) -1}\,
\exp\left({\frac{E_{\rm u}^2 -E_{\rm u}^1}{kT}}\right)
\label{eq:tau-r}
\end{equation}
For $| E_{\rm u}^1 -E_{\rm u}^2| << kT$ the exponential factor is
close to unity.

In our data there are pairs of transitions with similar upper state energies.
The $21_{0,21}-20_{0,20}$ and $10_{2,9/8}-9_{2,8/7}$ transitions have
similar $E_{\rm u}\sim 200$~K. However, the first one has higher
line strength and higher transition frequency; therefore, according to 
Eq.~(\ref{eq:tau-r}) it should have higher optical depth than the second
one. Then, it should be stronger influenced by possible
optical depth effects and the corresponding point in
Fig.~\ref{fig:ori-rd} should lie {\it lower} than the point
corresponding to the $10_{2,9/8}-9_{2,8/7}$ transitions. 
However, this is not a case. 
Actually, the points are very close to each other and perhaps
slightly shifted in the opposite sense. 
The same is true for the $21_{2,20/19}-20_{2,19/18}$ and 
$10_{3,8/7}-9_{3,7/6}$ transitions with $E_{\rm u}\sim 400$~K.
We conclude that the optical
depth for Orion in these transitions should be low. 
Perhaps in some other transitions or in other sources optical depths are as
high as indicated by our tentative HN$^{13}$CO detection. There is however
no reason to apply optical depth corrections to the bulk of
our sources.

\begin{table}
\caption[]{HNCO rotational temperatures, column densities and 
relative abundances.}
\begin{tabular}{llll}
\hline\noalign{\smallskip}
Source	&$T_{\rm rot}$	&$\log N_{\rm L}$	&$\log \chi$(HNCO)\\
&(K)	&(cm$^{-2}$)	&\\
\noalign{\smallskip}\hline\noalign{\smallskip}
Orion A		&25	&14.87	&--8.06\\
		&150	&15.00\\
		&530	&14.61\\
G 301.12	&24	&13.73	&--9.16\\
		&76	&13.94\\
G 305.20	&102	&13.69	&--9.33\\
G 308.80	&236	&13.89	&--8.99\\
G 329.03	&60	&13.89	&--8.91\\
G 330.88	&133	&13.96	&--9.34\\
G 332.83	&98	&13.94	&--9.65\\
G 337.40	&75	&13.72	&--9.58\\
G 340.06	&130	&13.64	&--9.06\\
		&550	&13.98\\
G 345.00	&88	&13.93	&--9.08\\
G 351.41	&93	&13.93	&--9.17\\
		&320	&13.96\\
G 351.58	&70	&13.65	&--9.39\\
G 351.78	&155	&14.64	&--8.88\\
Sgr A		&12	&14.95	&--8.20\\
S 158		&28	&14.03	&--9.23\\
S 255		&170	&13.68	&--9.17\\
W 49 N		&100	&15.00	&--8.58\\
W 51 M		&9	&14.44	&--8.74\\
		&38	&14.84\\
		&73	&14.49\\
W 75 N		&46	&14.38	&--8.92\\
W 75(OH)	&37	&14.38	&--9.07\\
\noalign{\smallskip}\hline
\end{tabular}
\label{table:rot}
\end{table}

Transitions with low $E_{\rm u}/k$ values
are fitted by rather low temperature models, 
$T_{\rm R}\sim 10-30$~K. Transitions between higher excited states
are related to higher
rotational temperatures up to $T_{\rm rot}\sim 500$~K. 
In Table~\ref{table:rot} we also present estimates of the HNCO
relative abundances. The hydrogen column densities have been
calculated from the C$^{18}$O data under the assumptions of LTE and
a C$^{18}$O relative abundance of $1.7\,10^{-7}$ (Frerking et
al. 1982). 
Typical HNCO
abundances are $\sim 10^{-9}$. Sgr~A does not look very exceptional here.
The relative HNCO abundance in Sgr~A is about the same as in Orion but
the rotational temperature is much lower. 
In contrast to many other sources there is no high { excitation} temperature
component in Sgr~A, indicating that the dense gas is { probably} cool. 
This agrees
with results from H\"uttemeister et al. (1998) based on SiO and C$^{18}$O.
The opposite scenario, a hot highly subthermally excited low density gas 
component ($n$(H$_2$) $\sim 10^4$ cm$^{-3}$)
as observed by H\"uttemeister et al. (1993) in 
ammonia toward Sgr B2 is less likely, due to the correlations 
between HNCO and SiO that will be outlined in Sects.~\ref{sec:compar} and 
\ref{sec:chem}.

It is important to emphasize that our estimates give lower limits 
to the relative abundance $X$(HNCO) = $N$(HNCO)/$N$(H$_2$)
for at least two reasons. First, the HNCO sources are much
more compact than their C$^{18}$O counterparts and tend to be spatially
unresolved. Our estimates give   
beam averaged values and ``real'' abundances in regions of HNCO
line formation should be significantly higher. Second, if the HNCO optical
depth is high we would underestimate its
column densities.

Next, we have to mention that all these estimates refer to the bulk of
the cores. In the high velocity gas the HNCO abundances are apparently
much higher. 

One might think that better estimates of HNCO abundances can be
obtained from comparison with the dust emission rather than with
C$^{18}$O. As shown, HNCO probably arises in ``warm'' environments
and in the dust emission we see preferentially a high temperature
medium while in  C$^{18}$O the reverse is true. However, interferometric
observations in Orion (\cite{blake96}) show that HNCO $K_{-1}=2$ and
dust distributions do not entirely coincide.
At the same time, as shown in
Sect.~\ref{sec:ir}, there is a tight correlation between the FIR emission
at 100~$\mu$m and C$^{18}$O(2--1) integrated line intensity. Therefore,
no large differences between estimates of HNCO abundances by both methods
can be expected. There are detailed studies of dust emission towards
some of our sources with comparable angular resolution. 
E.g. Henning et al. (2000)
show that total gas column densities derived from dust and from C$^{18}$O(2--1)
in G301.12--0.20 coincide within a factor of 3.

\begin{figure}
\resizebox{\hsize}{!}{\rotatebox{-90}{\includegraphics{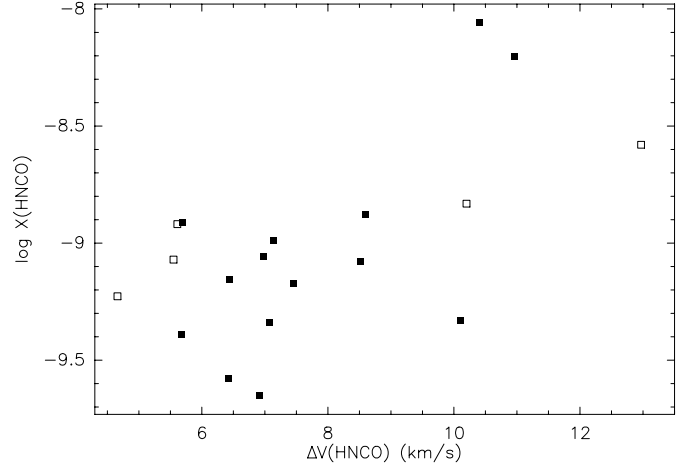}}}
\caption[]{HNCO relative abundance versus the HNCO line width for 
sources observed at SEST (filled squares) and in Onsala (open squares)}
\label{fig:r-hnco-c18o-dv}
\end{figure}

In Fig.~\ref{fig:r-hnco-c18o-dv} we plot the HNCO abundances versus
the HNCO line widths. There is a trend of increasing the
HNCO abundance with increasing HNCO line width. This shows that the
HNCO production can be related to dynamical activity in the sources.

Table~\ref{table:rot} and Fig.~\ref{fig:r-hnco-c18o-dv} indicate 
that abundances derived for the sources which belong 
to the inner and to the outer Galaxy, respectively, are about the same.
Therefore, there is no significant galactic gradient in HNCO abundance.  

\subsection{Physical conditions in regions of HNCO emission}

Now we shall try to understand the physical conditions in regions of
HNCO emission detected by us. An important question to start with is
which excitation mechanism dominates, radiative or collisional? And
which gas parameters are implied by each of them? To
answer these questions properly would require a numerical model taking both
into account. Useful conclusions can, however, also be obtained by
semi-qualitative consideration presented below. 
We concentrate here on Orion KL as the best studied source.

At first, we need an estimate for the size of the HNCO emission region.
Our map presented in Fig.~\ref{fig:maps} gives an upper limit of 
$\sim 10\arcsec$ for the $21_{0,21}-20_{0,20}$ transition. Interferometric
results (\cite{blake96}) give a size of $\sim 2\arcsec\times 4\arcsec$
for the $K_{-1}=2$ transition at 220 GHz. This can be probably considered
as an upper limit also for higher $K_{-1}$ ladders. On the other hand we
can obtain a lower limit on the source size from the comparison of the
brightness and excitation temperatures. For $T_{\rm ex}\la 500$~K (as
follows from the population diagram) we obtain that 
the lower limit on the beam filling factor for the
$K_{-1}=5$ transitions in Orion is $\sim 2\,10^{-3}$. Therefore, the effective
size of the emitting region is $\ga 1$\arcsec\ or $\ga 0.002$~pc,
i.e. $\ga 7\,10^{15}$~cm.

Let us consider the physical requirements in the case of
collisional excitation. The critical densities defined as 
$n_{\rm c}=A_{\rm ul}/C_{\rm ul}$ ($A_{\rm ul}$ is the spontaneous
decay rate and $C_{\rm ul}$ is the collisional de-excitation rate; 
Scoville et al. 1980) are
$\sim 10^6$~cm$^{-3}$ for the $10_{0,10}-9_{0,9}$ transition and
$\sim 10^7$~cm$^{-3}$ for the $21_{0,21}-20_{0,20}$ transition 
({ the collisional rates are $\sim 10^{-10}$~s$^{-1}$cm$^{-3}$}
as obtained from Sheldon Green's program
available on Internet -- http://www.giss.nasa.gov/data/mcrates/).
Much higher densities are needed for excitation of the transitions in
the $K_{-1}>0$ ladders. This is caused by fast $b$-type transitions
between different $K_{-1}$ ladders. E.g. the spontaneous emission rate
from the $K_{-1}=5$ ladder to the $K_{-1}=4$ ladder is $\sim
5$~s$^{-1}$. This implies a critical density of $\sim
10^{11}$~cm$^{-3}$. The gas kinetic temperature should be $\ga
500$~K. 

Such conditions cannot be excluded. Walker et al. (1994) derived
from observations of vibrationally excited CS $n\ga
10^{11}-10^{12}$~cm$^{-3}$ and $T\ga 1000$~K in a region $\sim
10^{15}$~cm from the stellar core
toward IRAS 16293--2422.
The question is whether the
required amount of such gas is consistent with the observations.

Taking into account the lower limit on the source size
the mass of the hot dense gas ($n\sim
10^{11}$~cm$^{-3}$, $T\ga 500$~K) would be $\ga
100$~M$_\odot$. Estimates of the hot core mass from dust continuum
measurements give values of $\sim 5-40$~M$_\odot$
(Masson \& Mundy 1988, Wright et al. 1992). Taking into account
the uncertainties in our estimations we cannot entirely exclude the possibility
of collisional excitation even for the $K_{-1}=5$ ladder but this appears
to be an unlikely scenario.

For the lower $K_{-1}$ ladders the density requirements can be
significantly relaxed. E.g. for the $b$-type transitions from the
$K_{-1}=3$ to the $K_{-1}=2$ ladder the spontaneous decay rate is $\sim
1$~s$^{-1}$ and the critical density is $\sim 10^{10}$~cm$^{-3}$.

The transitions in the $K_{-1}=0$ ladder, of course, will be also
excited in this hot dense gas. However, the emission in these lines
will be dominated by a more extended lower density component.

Now let us turn to radiative excitation. It requires
sufficient photons at the wavelengths
corresponding to the $b$-type transitions between different $K_{-1}$
ladders, from $\sim 300$ to $\sim 30$~$\mu$m. If the dilution factor
is close to unity we need an optical depth $\tau \ga 1$ and a
radiation temperature $T_{\rm R}\ga 500$~K at least at 30~$\mu$m. 
As an upper limit to the
source size we can take the { mean interferometric} value of $\sim
3$\arcsec. However, what will be the IR flux and 
luminosity of such a source? For the flux at 30~$\mu$m we obtain
$F(30\mu{\rm m})\approx 3\,10^4(\theta_{\rm s}/1\arcsec)^2$~Jy. The 
observational value is $\sim 5\,10^4$~Jy (van Dishoeck et al. 1998). 
Therefore, the angular source size should be $\theta_{\rm s}\la 1{\farcs}5$
and the linear size $\la 10^{16}$~cm. 
This practically
coincides with the lower limit on the source size derived from the
beam dilution (see above). Taking the dust
absorption coefficient of $k_{\rm m}\sim 10^2$~cm$^2$/g (Ossenkopf \&
Henning 1994) we conclude that the gas density in this region should be
$n\ga 3\,10^7$~cm$^{-3}$. In this case we have no problem to reconcile
the mass estimates with the available data.

However, at longer wavelengths the IR pumping from such a source might be
not sufficient. Say, for $\tau\propto \lambda^{-2}$ the optical depth at
300~$\mu$m will be only $\sim 0.01$. Therefore, we need even higher
gas column and volume densities and/or larger source sizes at longer 
wavelengths. The latter implies the presence of a 
temperature gradient in the source 
which is natural for an internally heated object.
The lower $K_{-1}$ ladders are apparently excited by radiation with a
lower effective temperature. 

To conclude, it is much easier to explain the
excitation of the higher $K_{-1}$ ladders by the radiative
process. The source size in Orion should be $1\arcsec -2\arcsec$
which agrees with the interferometric image in the $K_{-1}=2$
transition at 1.3~mm (Blake et al. 1996). 

The emission in the $K_{-1}=0$ ladder should be more extended. For
Orion again from a comparison between the brightness and excitation
temperatures the source size should be $\ga 6\arcsec$. 
Such a large source size for the $K_{-1} = 0$ transitions implies that 
the radiative excitation via $K_{-1} > 0$ ladders will become
inefficient. Therefore, for the $K_{-1} = 0$ ladder collisional excitation
may dominate which implies gas densities $n\ga 10^6-10^7$~cm$^{-3}$. 
This scenario
is supported by several sources where the HNCO emission peak is
significantly displaced from any known IR source. 
The most obvious example is G~270.26+0.83 (Fig.~\ref{fig:maps}).
This implies either the presence of a very dense prestellar core or a
highly obscured young stellar object at this location.


\subsection{Comparison with C$^{18}$O, CS and SiO data}
\label{sec:compar}

An obvious step ahead to understand the properties of interstellar HNCO
emission is to compare our results with data from other better studied species.
The most reliable comparison can be done with our C$^{18}$O data 
which were observed simultaneously with
HNCO.

\begin{figure}
\resizebox{\hsize}{!}{\includegraphics{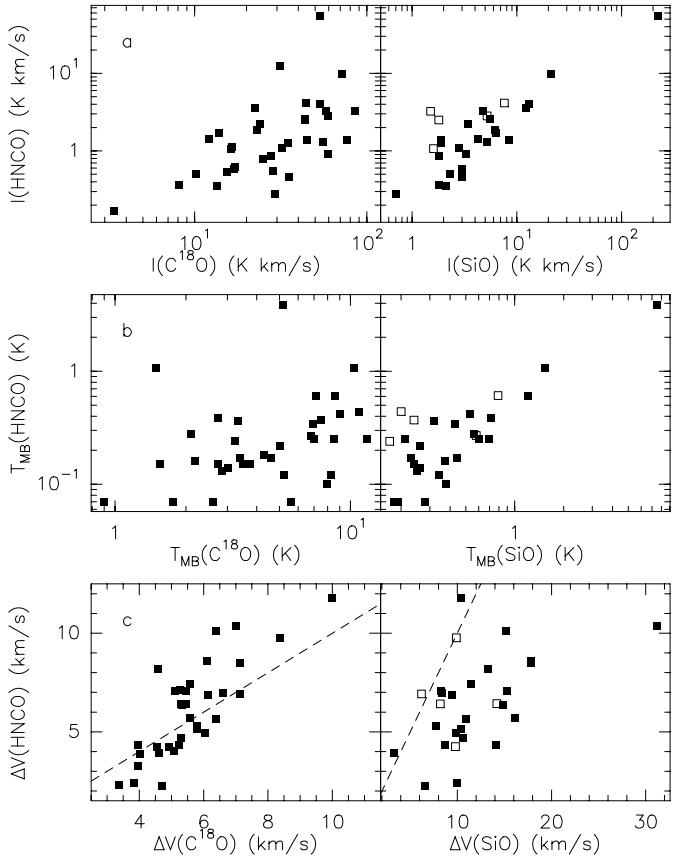}}
\caption[]{HNCO $10_{0,10}-9_{0,9}$
integrated line intensities, peak main beam temperatures 
and line widths versus corresponding C$^{18}$O $J=2-1$ 
and SiO $J=3-2$ peak temperatures
for the SEST sample. Open squares correspond to those SiO data which were
obtained at slightly different positions than HNCO. The dashed lines in the
panel (c) correspond to equal line widths of the compared species}
\label{fig:compar}
\end{figure}

Fig.~\ref{fig:compar} shows a noticeable correlation between the HNCO and
C$^{18}$O integrated line intensities. However, it is produced apparently
by the correlation between the line widths since the correlation between 
HNCO and C$^{18}$O peak line temperatures is rather weak.

The plot of $\Delta V(\mbox{HNCO})$ versus $\Delta V(\mbox{C$^{18}$O})$
looks rather interesting. 
Concerning the 220~GHz transitions for the narrowest C$^{18}$O lines
the HNCO line width is smaller than that of C$^{18}$O. With increasing
C$^{18}$O linewidth, however, the HNCO lines broaden faster and become
broader than the C$^{18}$O lines. An exception is Sgr~A (not shown in the 
plot) but its C$^{18}$O
spectrum is strongly distorted by emission from the reference
position. 

A similar comparison with the CS(2--1) data from Zinchenko et al. (1995,
1998) and Juvela (1996) (not shown here)
shows even lower correlations between the line parameters than in the case
of C$^{18}$O. However, in this case the beam sizes for
CS and HNCO are different and even the central positions not always
coincide.

In contrast, much better correlations exist between the HNCO and SiO line
parameters (the latter ones are taken from Harju et al. 1998). 
Good correlations exist for both integrated and peak intensities. 
The correlation between the line widths
is somewhat worse  
but one should take into account that the SiO line widths
were derived from the second moments of the line profiles while the HNCO
widths represent results of the gaussian fits. Anyway, the correlation
does exist and the SiO lines are almost always broader than the HNCO lines.

\begin{figure}
\resizebox{\hsize}{!}{\rotatebox{-90}{\includegraphics{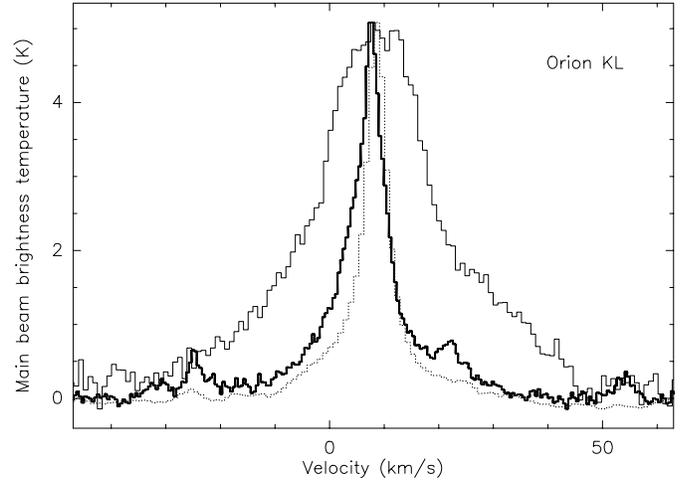}}}
\caption[]{The HNCO $10_{0,10}-9_{0,9}$ line in Orion (thick line) in
comparison with the C$^{18}$O(2--1) (dotted line) and SiO(2--1) (thin solid
line). The latter ones are scaled to the same peak intensity as HNCO
$10_{0,10}-9_{0,9}$ }
\label{fig:ori-over}
\end{figure}

A more detailed comparison with other species should include the
line profiles. For Orion, such a comparison 
is displayed in Fig.~\ref{fig:ori-over}.
It shows that HNCO lines possess an extra wing emission which is  
less pronounced than in SiO. A similar picture is seen in some other
sources.

This comparison shows that HNCO is closely related to SiO
which is thought to be produced primarily in shocks and other energetic
processes. 
The comparison with the presumably optically thin C$^{18}$O(2--1) line
shows that the HNCO/CO abundance ratio is apparently enhanced in high velocity 
gas although to a lower degree than for SiO. Since the CO abundance is
usually assumed to be constant in bipolar flows (e.g., \cite{cabrit92};
\cite{shepherd96}) 
we see that HNCO abundances are enhanced relative to hydrogen, too.

It is interesting to note that the interferometric data for Orion
(\cite{blake96}) show that the spatial distributions of SiO and HNCO are
rather different. However, this does not exclude a common production
mechanism. E.g. these species can be formed at different stages in the
postshock gas.

\subsection{Comparison with IR data}\label{sec:ir}

\begin{figure}
\resizebox{\hsize}{!}{\rotatebox{-90}{\includegraphics{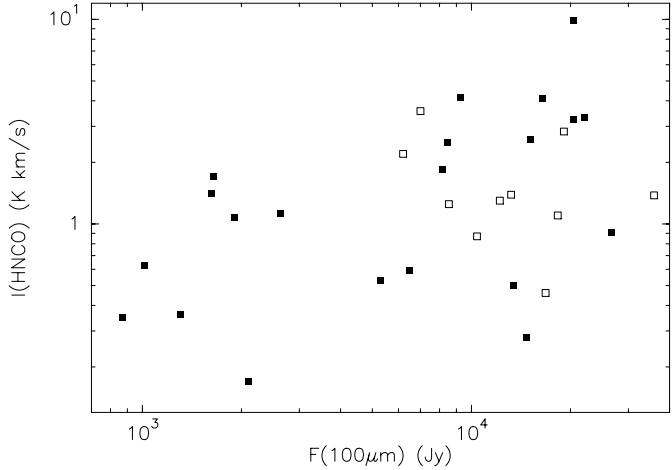}}}
\caption[]{HNCO $10_{0,10}-9_{0,9}$
integrated line intensities versus the FIR flux at 100~$\mu$m for the 
SEST sample. The open squares correspond to the cases where there is a large
($\ga$ HPBW) displacement between the position observed in HNCO and the
IRAS position}
\label{fig:hnco-ir}
\end{figure}

\begin{figure}
\resizebox{\hsize}{!}{\rotatebox{-90}{\includegraphics{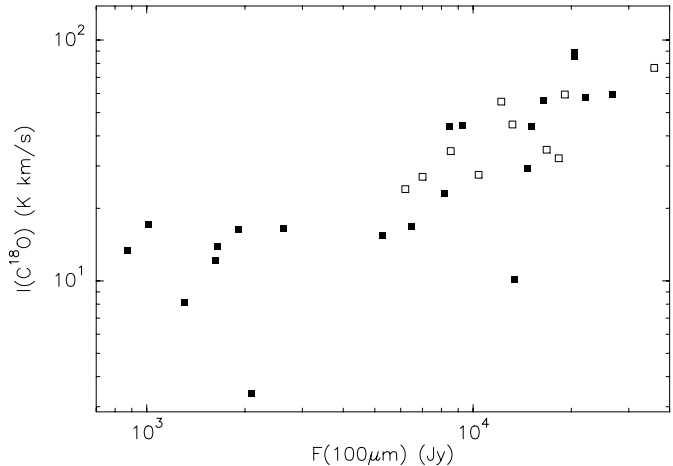}}}
\caption[]{C$^{18}$O(2--1)
integrated line intensities versus the FIR flux at 100~$\mu$m for the 
SEST sample. The open squares correspond to the cases where there is a large
($\ga$ HPBW) displacement between the position observed in C$^{18}$O and the
IRAS position}
\label{fig:c18o-ir}
\end{figure}

The correlation between HNCO integrated line intensities and FIR flux,
e.g. at 100~$\mu$m taken from IRAS data
(Fig.~\ref{fig:hnco-ir}), looks rather similar to the 
relationship between HNCO and C$^{18}$O (Fig.~\ref{fig:compar}). This is
natural because there is a rather tight correlation between the 100~$\mu$m
flux and the C$^{18}$O integrated line intensity (Fig.~\ref{fig:c18o-ir}). 
Such a good correlation shows that C$^{18}$O relative abundances 
are rather constant and justifies the usage of the HNCO/C$^{18}$O
ratio for estimation of HNCO abundances.

\subsection{HNCO chemistry}
\label{sec:chem}

In the early work of Iglesias (1977) HNCO was suggested to form via
ion-molecule reactions.  The sequence leading to HNCO via electron 
recombination of
H$_2$NCO$^+$ is initiated by the
formation of NCO$^+$ (either by a reaction between CN and O$_2^+$
or between He$^+$ and NCO; see also \cite{brown81}). 
The predicted HNCO abundances from
this reaction scheme are low. The steady state fractional abundance is of
the order $10^{-10}$ for a model with $n_{\rm H_2} \sim 10^4$ cm$^{-3}$
(\cite{iglesias}), and still lower for higher densities, because the
fractional ion abundances are roughly inversely proportional to the square
root of the gas density.

The abundances derived from ion-molecule chemistry are in contradiction with
the observations, especially when HNCO is believed to trace high density
gas. Recently, a new neutral gas-phase pathway has been suggested by Turner
et al. (1999) for translucent clouds: 
${\rm CN} + {\rm O_2} \rightarrow {\rm NCO} + {\rm O}$
followed by ${\rm NCO} + {\rm H_2} \rightarrow {\rm HNCO} + {\rm H}$. The
importance of these reactions can however be questioned, since 1) the
abundance of O$_2$ in the interstellar space is poorly known; and 2) the
second reaction probably has an activation barrier of about 1000 K
(\cite{turner99}).

Chemistry models predict high fractional O$_2$ abundances (up to $\sim
10^{-5}$) at late stages of chemical evolution in dense cores and in
postshock gas (e.g \cite{caselli93}; \cite{bergin98}). However, the upper
limits derived from observations towards several GMC cores (most recently by
the SWAS satellite; \cite{melnick}) are about $10^{-6}$, which indicates that 
the oxygen
chemistry is not well understood, yet. O$_2$ is destroyed by UV radiation
and in powerful shocks (with shock velocities greater than 26 kms$^{-1}$;
\cite{bergin98}), and is therefore likely thriving in relatively
quiescent dense gas or in regions associated with low velocity shocks. The
same should be true for HNCO if the reaction suggested by Turner et al.
(1999) is relevant. 

The observed correlation between SiO and HNCO integrated line intensities 
indicates the prevalence of shocks in the HNCO emission regions. Shock
heating can therefore provide the means of overcoming the energy barrier in the reaction
between NCO and H$_2$, and thereby intensify the HNCO production. 
On the other hand, the fact that the HNCO line widths are smaller than those of SiO  
could be understood by the destruction of O$_2$ in high velocity shocks. 

In the light of the present observations the neutral reactions suggested by
Turner et al. (1999) appear to provide a plausible production pathway of HNCO
also in warm GMC cores. The formation of HNCO via grain surface reactions,
e.g. through the desorption and subsequent fragmentation of some more
complex molecule is an alternative, which to our knowledge has not
yet been investigated.

\section{Conclusions}

We have presented the results of an HNCO survey of high mass star-forming cores
at frequencies from 22 to 461~GHz.
The main conclusions are the following:

\begin{enumerate}
\item HNCO is widespread in dense cores forming high mass stars.
The detection rate was $\sim 70$\%. 
There is no significant galactic gradient in its abundance
as indicated by the fact
that abundances derived for the sources which belong  
to the inner and to the outer Galaxy, respectively, are about the same.

\item Transitions in higher $K_{-1}$ ladders, up to $K_{-1}=5$, 
are detected. The excitation energy reaches $\sim 1300$~K above the
ground level.

\item HN$^{13}$CO is tentatively detected { towards G\,301.12--0.20}. 
This implies an optical depth
in the HNCO $10_{0,10}-9_{0,9}$ line $\sim 10$ 
in this source. The optical depth in the $1_{01}-0_{00}$ transition
is $\tau\la 1$ for the sources detected in this line as inferred from the
hyperfine ratios.

\item The sources are compact with sizes $\la 20\arcsec$.

\item HNCO rotational temperatures vary from $\sim 10$~K to $\sim 500$~K.
Typical relative abundances are $\sim 10^{-9}$. These increase with
increasing velocity dispersion.

\item The emission in the $K_{-1}>0$ ladders is best explained by
FIR radiative excitation. 
In order to provide a sufficiently large dust optical depth at FIR
wavelengths taking into account the limitations on the source size,
the gas density should be $n\ga 3\,10^7$~cm$^{-3}$;
a temperature  
$T\ga 500$~K is needed to excite the $K_{-1}=5$ emission in Orion KL.
The $K_{-1}=0$ transitions can be collisionally excited. The required
densities are $n\ga 10^6-10^7$~cm$^{-3}$.

\item HNCO correlates well with SiO and does not correlate with CS which 
is a typical high density probe. HNCO abundances are enhanced in high velocity
gas. Probably HNCO production is related to shocks as for SiO. A plausible
pathway is gas-phase neutral-neutral reactions at high ($> 1000$ K) 
temperatures to overcome an activation barrier that is likely inhibiting
the ${\rm NCO} + {\rm H_2} \rightarrow {\rm HNCO} + {\rm H}$ reaction in
a cool interstellar medium.

\end{enumerate}

\begin{acknowledgements}
We are very grateful to Dr. J.~Harju for his contribution to this work, to
Dr. Lars E.B. Johansson for the help with the observations
in Onsala, to the SEST staff, 
to Alexander Lapinov for calculating HNCO line strengths
and to the referee, Dr. C.M.~Walmsley, for the very useful detailed 
comments. 
I.Z. thanks
the Helsinki University Observatory and Max-Planck-Institut f\"ur
Radioastronomie for the hospitality.
He was also supported in part by the DFG grant 436 RUS 113/203/0,
INTAS grant 93-2168-ext, NASA grant
provided via CRDF RP0-841 and grants 
96-02-16472, 99-02-16556 from the Russian Foundation for Basic Research.
This research has made use of the Simbad database, operated at CDS,
Strasbourg, France. 
\end{acknowledgements}


\begin{thebibliography}{}

\bibitem[Baars \& Martin 1996]{baars96} 
Baars, J., Martin, R.N., 1996, Rev. Mod. Astron. 9, 111 
\bibitem[Bergin et al. 1998]{bergin98}Bergin E. A., Melnick G.J., Neufeld
D.A. 1998, ApJ 499, 777
\bibitem[Blake et al. 1996]{blake96} Blake G.A., Mundy L.G., Carlstrom J.E., 
et al., 1996, ApJ 472, L49
\bibitem[Brown 1981]{brown81}Brown R.L. 1981, ApJ 248, L119
\bibitem[Cabrit \& Bertout 1992]{cabrit92}
Cabrit S., Bertout C., 1992, A\&A 261, 274
\bibitem[Caselli et al. 1993]{caselli93}
Caselli P., Hasegawa T.I., Herbst E. 1993, ApJ 408, 548
\bibitem[Churchwell et al. 1986]{churchwell86} Churchwell E., Wood D., 
Myers P.C., Myers R.V. 1986, ApJ 305, 405
\bibitem[Dahmen et al. 1997]{dahmen97} Dahmen G., H\"uttemeister S., 
Wilson T.L., Mauersberger R., 1997, A\&AS 126, 197
\bibitem[Downes 1989]{downes89} Downes, D., 1989, Evolution of Galaxies -- 
Astronomical Observations,
        Lecture Notes in Physics 333, eds. I. Appenzeller, H. Habing,
        P. L{\'e}na, Springer Verlag, Berlin, p353 
\bibitem[Frerking et al. (1982)]{frerking82}
Frerking M.A., Langer W.D., Wilson R.W., 1982, ApJ 262, 590
\bibitem[Goldsmith \& Langer 1999]{goldsmith99}Goldsmith P.F., Langer W.D.,
1999, ApJ 517, 209
\bibitem[Groesbeck et al. 1994]{groesbeck94} Groesbeck T.D., Phillips T.G., 
Blake G.A., 1994, ApJS 94, 147
\bibitem[Harju et al. 1998]{harju98} 
Harju J., Lehtinen K., Booth R., Zinchenko I., 1998, A\&AS 132,
211
\bibitem[Harris et al. 1995]{harris95}Harris A.I., Avery L.W., Schuster K.-F.,
Tacconi L.J., Genzel R., 1995, ApJ 446, L85
\bibitem[Henning et al. 2000]{henning00}
Henning Th., Lapinov A., Schreyer K., Stecklum B., Zinchenko I., 2000,
A\&A, submitted
\bibitem[H\"uttemeister et al. 1999]{huttemeister93}
H\"uttemeister S., Wilson T. L., Henkel C., Mauersberger R., 1993,
A\&A 276, 445
\bibitem[H\"uttemeister et al. 1998]{huttemeister98}
H\"uttemeister S., Dahmen G., Mauersberger R., et al., 1998, A\&A 334, 646
\bibitem[Iglesias 1977]{iglesias}Iglesias E. 1977, ApJ 218, 697
\bibitem[Jackson et al. (1984)]{jackson84} Jackson J.M., Armstrong J.T., 
Barrett A.H. 1984, ApJ 280, 608
\bibitem[Juvela 1996]{juvela96} Juvela M., 1996, A\&AS 118, 191
\bibitem[Kuan \& Snyder 1996]{kuan96} Kuan Y.-J., Snyder L.E. 1996,
ApJ 470, 981
\bibitem[Lapinov et al. 1998]{lapinov98} 
Lapinov A.V., Schilke P., Juvela M., Zinchenko I., 1998, A\&A 336, 1007
\bibitem[Lindqvist et al. 1995]{lindqvist95} Lindqvist M., Sandqvist Aa.,
Winnberg A., Johansson L.E.B., Nyman L.-\AA. 1995, A\&AS 113, 257
\bibitem[Mart\'{\i}n-Pintado et al. 1990]{martin-pintado90}
Mart\'{\i}n-Pintado J., de Vicente P., Wilson T.L., Johnston K.J., 1990,
A\&A 236, 193
\bibitem[Masson \& Mundy 1988]{masson88} Masson C.R., Mundy L.G., 1988,
ApJ 324, 538
\bibitem[Melnick et al. 1999]{melnick}Melnick G.J., Stauffer J.R., Ashby
M.L.N, et al. 1999, BAAS 194, 4708
\bibitem[Ossenkopf \& Henning 1994]{ossenkopf94}Ossenkopf V., Henning Th.,
1994, A\&A 291, 943
\bibitem[Ott et al. 1994]{ott94} 
Ott M., Witzel A., Quirrenbach A., et al., 1994, A\&A 284, 331
\bibitem[Sato et al. 1997]{sato97} Sato F., Hasegawa T., Whiteoak J.B., 
Shimizu M., 1997, 
In: IAU  Symposium No. 184. ``The Central Regions of the Galaxy 
                    and Galaxies''.  Kyoto, Japan, 17-30 August, 1997, p.~98
\bibitem[Schulz et al. 1995]{schulz95} 
Schulz, A., Henkel, C., Beckmann, U., et al., 1995, A\&A 218, 24 
\bibitem[Schilke et al. 1997]{schilke97} Schilke P., Groesbeck T.D., 
Blake G.A., Phillips T.G., 1997, ApJS 108, 301
\bibitem[Scoville et al. 1980]{scoville80}
Scoville N.Z., Krotkov R., Wang D., 1980, ApJ 240, 929
\bibitem[Shepherd \& Churchwell E. 1996]{shepherd96} 
Shepherd D.S., Churchwell E., 1996, ApJ 472, 225
\bibitem[Snyder \& Buhl (1972)]{snyder72} Snyder L.E., Buhl D., 1972,
ApJ 177, 619
\bibitem[Sutton et al. 1995]{sutton85} Sutton E.C., Blake G.A., Masson C.R.,
Phillips T.G., 1985, ApJS 58, 341
\bibitem[Townes \& Schawlow]{townes} 
Townes C.H., Schawlow A.L. 1975, Microwave Spectroscopy, Dover
Publications, New York 
\bibitem[Turner et al. 1999]{turner99}Turner B.E., Terzieva R., Herbst E.
1999, ApJ 518, 699
\bibitem[van Dishoeck et al. 1998]{dishoeck98} van Dishoeck E.F., Wright C.M.,
Cernicharo J., et al., 1998, ApJ 502, L173
\bibitem[Walker et al. 1994]{walker94} 
Walker C.K., Maloney P.R., Serabyn E., 1994, ApJ 437, L127
\bibitem[Walsh et al. 1998]{walsh98} 
Walsh A.J., Burton M.G., Hyland A.R., Robinson G., 1998,
MNRAS 301, 640
\bibitem[Wilson et al. 1996]{wilson96} Wilson T.L., Snyder L.E., Comoretto G., 
Jewell P.R., Henkel C., 1996, A\&A 314, 909
\bibitem[Winnewisser et al. 1976]{winnewisser76} Winnewisser G., Hocking W.H.,
Gerry M.C.L., 1976, J. Chem. Phys. Ref. Data 5, 79
\bibitem[Wright et al. 1992]{write92} Wright M., Sandell G., Wilner D.J.,
Plambeck R.L., 1992, 393, 225
\bibitem[Zinchenko 1995]{zin95} 
Zinchenko I., 1995, A\&A 303, 554
\bibitem[Zinchenko et al. 1995]{zinetal95} 
Zinchenko I., Mattila K., Toriseva M., 1995, A\&AS 111, 95 
\bibitem[Zinchenko et al. 1998]{zin98} 
Zinchenko I., Pirogov L., Toriseva M., 1998, A\&AS 133, 337

\end{thebibliography}
\end{document}